\documentclass[structabstract]{aa}  
\usepackage{graphicx}
\usepackage{natbib}
\bibpunct{(}{)}{;}{a}{}{,} % to follow the A&A style
\usepackage{txfonts}
\usepackage{lscape}
\usepackage{longtable,lscape}
\usepackage{color}
\begin{document}
   \title{Proper motions of young stars in Chamaeleon}

   \subtitle{II. New kinematical candidate members of Chamaeleon~I and II}

 \author{Bel\'en L\'opez Mart\'{\i}
          \inst{1}
          \and
          Francisco Jim\'enez-Esteban \inst{1,2,3}
	  \and
	  Amelia Bayo \inst{4,5}
          \and
          David Barrado \inst{1,6}
	  \and 
	  Enrique Solano \inst{1,2}
          \and
          Herv\'e Bouy \inst{1}
	  \and
	  Carlos Rodrigo \inst{1,2}
          }

   \institute{
        Centro de Astrobiolog\'{\i}a (INTA-CSIC), Departamento de Astrof\'{\i}sica, P.O. Box 78, E-28261 Villanueva de la Ca\~nada, Madrid, Spain\\
	\email{belen@cab.inta-csic.es}
	\and
	Spanish Virtual Observatory, Spain
	\and
	Suffolk University, Madrid Campus, C/ Valle de la Vi\~na 3, E-28003 Madrid, Spain
	\and
	European Southern Observatory, Alonso de C\'ordova 3107, Vitacura, Santiago, Chile 
	\and
	Max-Planck-Institut f\"ur Astronomie, K\"onigstuhl 17, D-69117 Heidelberg, Germany
	\and
	Calar Alto Observatory, Centro Astron\'omico Hispano-Alem\'an, C/ Jes\'us Durb\'an Rem\'on 2-2, E-04004 Almer\'{\i}a, Spain
	}

   \date{Received; accepted }
   
   \titlerunning{New kinematical candidate members of Chamaeleon}

% \abstract{}{}{}{}{} 
% 5 {} token are mandatory
 
  \abstract
   {
   The Chamaeleon star-forming region has been extensively studied in the last decades. However, most studies have been confined to the densest parts of the clouds. In a previous paper, we analysed the kinematical properties of the spectroscopically confirmed population of the Chamaeleon~I and II clouds.
    }
   {
   We want to search for new kinematical candidate members to the Chamaeleon~I and II moving groups, extending the studied area beyond the clouds, and to characterize these new populations using available information from public databases and catalogues. We also want to check if the populations of the moving groups are confined to the present dark clouds.
  }
  {
  Kinematic candidate members were initially selected on the basis of proper motions and colours using the Fourth US Naval Observatory CCD Astrograph Catalog (UCAC4). The SEDs of the objects were constructed using photometry retrieved from the Virtual Observatory and other resources, and fitted to models of stellar photospheres to derive effective temperatures, gravity values, and luminosities. Masses and ages were estimated by comparison with theoretical evolutionary tracks in a Hertzprung-Russell diagram. Objects with ages $\lesssim$20~Myr were selected as probable members of the moving groups.
   }
   {
   We have identified 51 and 14 candidate members to the Chamaeleon~I and II moving groups, respectively,  of which 17 and 1, respectively, are classified as probable young stars according to the SED analysis. Another object in Chamaeleon~I located slightly above the 1~Myr isochrone is classified as a possible young star.  All these objects are diskless stars with masses in the range $0.3\lesssim M/M_{\odot}\lesssim1.4$ and ages consistent with those reported for the corresponding confirmed members. They tend to be located at the boundaries of or outside the dark clouds, preferably to the north-east and south-east in the case of Chamaeleon~I, and to the north-east in the case of Chamaeleon~II.
    }
 {
 The kinematical population of Chamaeleon~I and II could be larger and spread over a larger area of the sky than suggested by previous studies. However, the results of this study should be confirmed with spectroscopic data and more precise kinematic information.
}

   \keywords{stars:low-mass, brown dwarfs $-$ stars: formation $-$ stars: pre-main sequence $-$ stars: luminosity function, mass function $-$ astronomical database: miscellaneous $-$ virtual observatory tools} 

   \maketitle

%$-$$-$$-$$-$$-$$-$$-$$-$$-$$-$$-$$-$$-$$-$$-$$-$$-$$-$$-$$-$$-$$-$$-$$-$$-$$-$$-$$-$$-$
   \begin{figure*}[t]
   \centering
       \includegraphics[width=\textwidth]{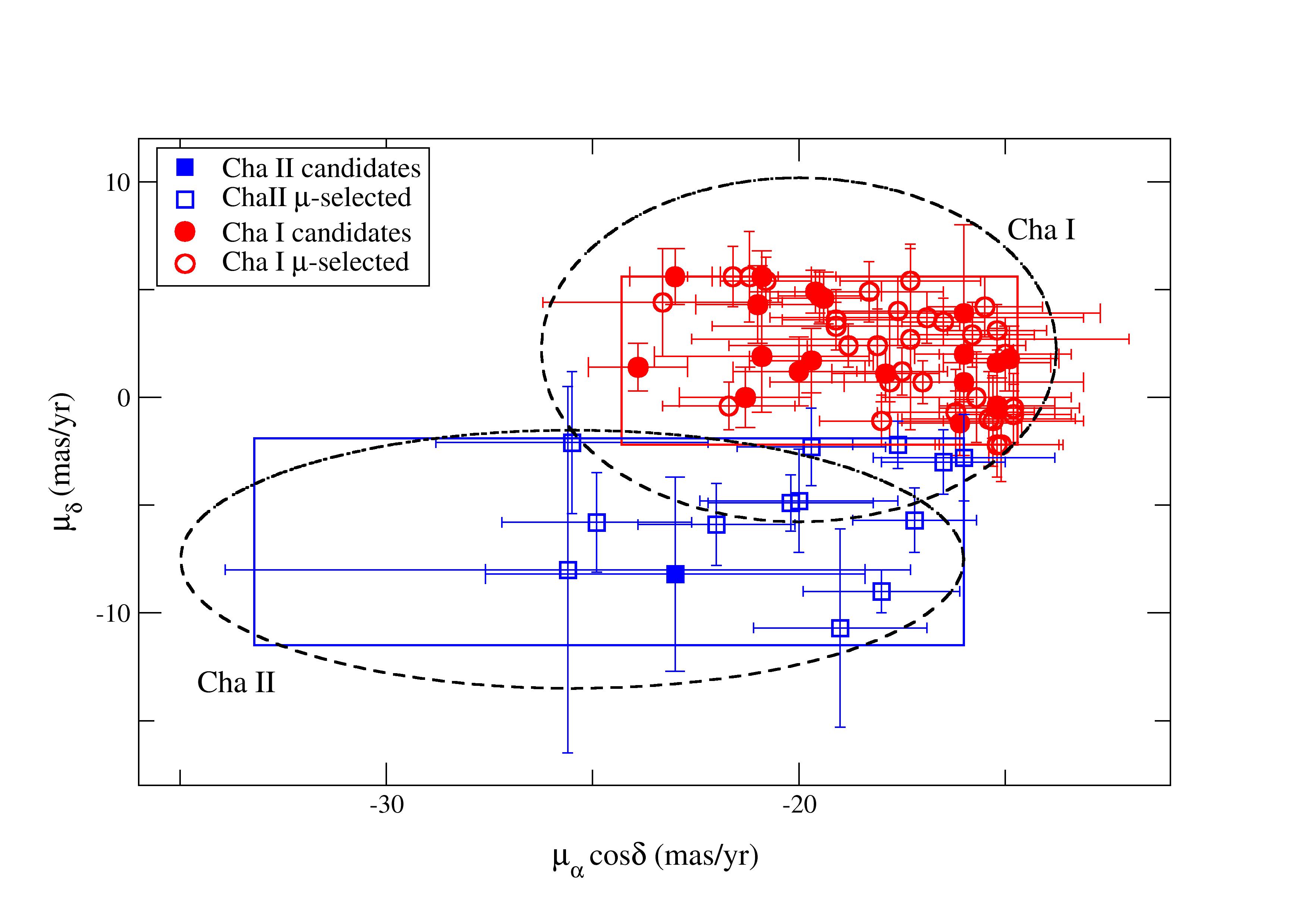}\hfill
\caption{\footnotesize
	       Vector point diagram for our kinematic candidate members of Chamaeleon~I and II (red circles and blue squares, respectively). The objects initially selected in Sect.~\ref{sec:cand} on the basis of proper motions and colours are marked with open symbols, and the final candidates from the refined selection based on the SED analysis from Sect.~\ref{sec:vosa} are shown with solid symbols. The ellipses indicate the approximate location of the kinematical groups identified in Paper~I. The boxes show the limits in proper motion for new candidate members defined in Table~\ref{tab:search}.  
	       }
         \label{fig:cand}
   \end{figure*}
%$-$$-$$-$$-$$-$$-$$-$$-$$-$$-$$-$$-$$-$$-$$-$$-$$-$$-$$-$$-$$-$$-$$-$$-$$-$$-$$-$$-$$-$

%
%########################################################################
\section{Introduction}\label{sec:intro}

The \object{Chamaeleon} complex is composed of three dark clouds, named Chamaeleon~I, II, and III \citep{schwartz1977}, lying at distances in the range 115-215~pc.  \object{Chamaeleon~I} (\object{Cha~I}), the oldest of the three ($\sim$2~Myr) contains more than 300 known young stars \citep[for the latest census, see][]{luhman2008a}, while  more than 60 stars have been identified in \object{Chamaeleon~II} \citep[\object{Cha~II};  $\lesssim$1~Myr;][and references therein]{spezzi2007,spezzi2008,alcala2008}. No young star population has been identified so far in \object{Chamaeleon~III} (Cha~III). 

Most studies of the stellar and substellar population in Chamaeleon have focused on the densest areas of the clouds and on the search for indications of youth such as infrared excesses, prominent H$\alpha$ emission, or accretion signatures in the object spectra \citep[e.g.][]{prusti1991,wfichai,wficrachaii,barrado2004,luhman2004,luhman2007,spezzi2007,alcala2008}. 
To assess the diskless component of the population, X-ray surveys have been carried out \citep[e.g.][]{feigelson1993,alcala2000,stelzer2004}. Thus, both the class II (with disk) and class III (diskless) populations have been identified in the core regions.

The areas outside the main cloud cores have not been studied as much. In the 1990s, the ROSAT All Sky Survey (RASS) unveiled a dispersed population of young X-ray emitting stars towards the Chamaeleon area, but not confined to the dark clouds \citep{alcala1995,alcala1997}. Later kinematical studies based on radial velocity \citep{covino1997} and proper motion measurements \citep{frink1998} showed that the sample included objects belonging to different kinematical populations, located at different distances, from 60 to about 200~pc, corresponding to the Cha~I cloud, the foreground $\epsilon$~Chamaeleontis ($\epsilon$~Cha) cluster, and an undetermined field population. Further interlopers from the $\epsilon$~Cha cluster were later identified in the Chamaeleon sky area  \citep[e.g.][]{mamajek1999,feigelson2003,luhman2004b,luhman2008c}. 

In a previous paper \citep[][hereafter Paper~I]{vocha1}, we used proper motions from the Third US Naval Observatory CCD Astrograph Catalog \citep[UCAC3;][]{zacharias2010} to study the moving groups seen towards the Chamaeleon sky area. We showed that the UCAC3 proper motions from the previously-known candidate members of the associations are precise enough to clearly distinguish the Cha~I and II populations from those of the foreground $\epsilon$~Cha and $\eta$~Chamaeleontis ($\eta$~Cha) clusters. Our study allowed us to identify a few interlopers among our compiled samples of confirmed young stars  in each association, and to provide further information on the membership of other candidates from the literature without spectroscopic confirmation of youth. In addition, our results suggested that the stars in Cha~I and II may constitute two different moving groups, which questions the physical relation between the clouds.

This paper is a continuation of the work presented in Paper~I, and reports on a search for new candidate members of the Chamaeleon clouds based on UCAC proper motions and colours, and taking advantage of Virtual Observatory (VO)\footnote{\footnotesize 
The VO is a project designed to provide the astronomical community with the data access and the research tools necessary to enable the exploration of the digital, multi-wavelength universe resident in the astronomical data archives. http://www.ivoa.net
} 
tools and protocols. Since UCAC is an all-sky catalogue, this approach allows us to explore a larger sky area than most of the previous searches reported in the literature. 

The study presented here was initially performed on UCAC3 data. However, during the publication process of Paper~I,  the fourth and last release of the UCAC catalogue (UCAC4) had become available \citep{zacharias2013}. The UCAC4 catalogue is an improved version of UCAC3, based on the same pixel reductions. Improvements over UCAC3 include bug fixes, correction of systematic errors, the inclusion of the brightest stars from the \emph{Hipparcos}, Tycho-2, and FK6 catalogues, and the use of APASS photometry in the Johnson B and V bands and the Sloan $g'$, $r'$, and $i'$ bands. It is thus a cleaner catalogue than UCAC3, but with very similar accuracy and coverage. We checked this by constructing the UCAC4 vector-point diagram for the confirmed members of the four associations, and comparing it with the UCAC3 diagram presented in Paper~I (Fig.~1 in that work). The main difference between the two diagrams is that the objects detached from the moving groups in the UCAC3 vector-point diagram, that were shown to have incorrect or dubious proper motion measurements, are missing from the UCAC4 diagram. Thus, the main conclusions from Paper~I are not altered when UCAC4 data are used instead of UCAC3. On the other hand, our comparison of both catalogues revealed a problem with the zeropoints of the SuperCosmos R and I photometry included in UCAC3, which made this photometry internally inconsistent. Since our candidate selection is based on colours as well as on proper motions (see Sect.~\ref{sec:sel}), we decided to use the latest version of the catalogue (UCAC4) and the more reliable APASS photometry included there for this part of our study. We note that no photometry from UCAC3 was used in the analysis of the confirmed members presented in Paper~I, and 
so this problem does not compromise the conclusions from that work.

%$-$$-$$-$$-$$-$$-$$-$$-$$-$$-$$-$$-$$-$$-$$-$$-$$-$$-$$-$$-$$-$$-$$-$$-$$-$$-$$-$$-$$-$
   \begin{figure*}[ht]
   \centering
    \includegraphics[width=18cm]{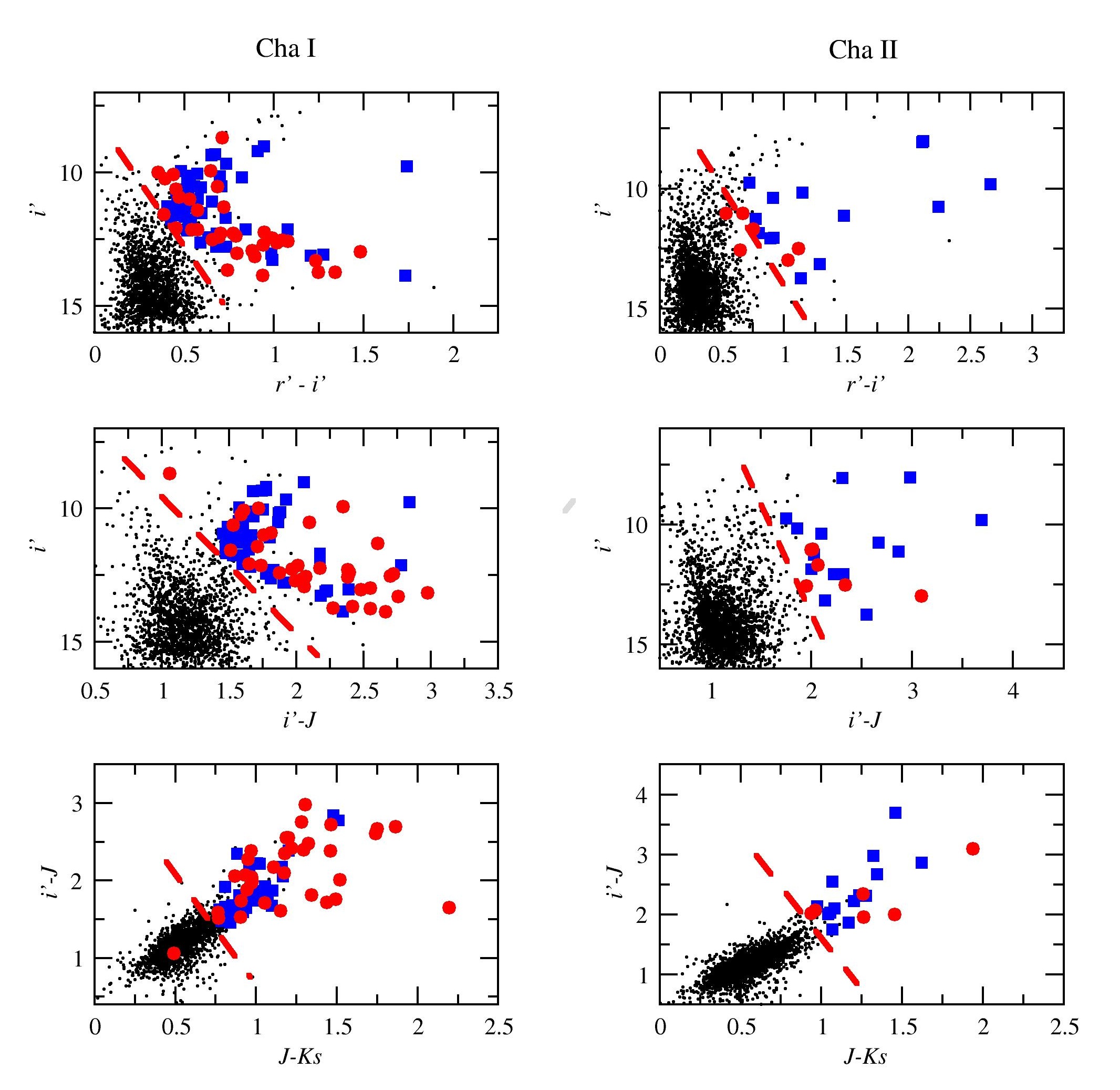}\hfill
       \caption{\footnotesize
Colour-colour and colour-magnitude diagrams used to select new candidate members of Cha~I  and II (blue squares) from the list of objects with compatible proper motions seen towards the same area of the sky (black dots). Only those objects that are located in all diagrams in the same area as most of the confirmed members (red circles) are selected. To help the eye, the thick dashed red lines indicate the approximate selection areas. See Sect.~\ref{sec:sel} for details.
	       }
         \label{fig:sel}
   \end{figure*}
%$-$$-$$-$$-$$-$$-$$-$$-$$-$$-$$-$$-$$-$$-$$-$$-$$-$$-$$-$$-$$-$$-$$-$$-$$-$$-$$-$$-$$-$

%$-$$-$$-$$-$$-$$-$$-$$-$$-$$-$$-$$-$$-$$-$$-$$-$$-$$-$$-$$-$$-$$-$$-$$-$$-$$-$$-$$-$$-$
%\addtocounter{table}{1} % For criteria
\begin{table*}[t]
\label{tab:search}
\caption{Astrometric selection criteria for new candidate members}
\centering
\begin{tabular}{l c c c c c}
\hline
\hline\noalign{\smallskip}
  \multicolumn{1}{l}{Cloud} &
   \multicolumn{1}{c}{$\alpha$ (J2000)} & 
    \multicolumn{1}{c}{$\delta$ (J2000)}  &
   \multicolumn{1}{c}{Radius} &
 \multicolumn{1}{c}{$\mu_{\alpha}\cos\delta$} &
  \multicolumn{1}{c}{$\mu_{\delta}$} \\
  \multicolumn{1}{l}{} &
   \multicolumn{1}{c}{hh mm ss} &
   \multicolumn{1}{c}{dd mm ss} &
   \multicolumn{1}{c}{(deg)} &
 \multicolumn{1}{c}{(mas/yr)} &
  \multicolumn{1}{c}{(mas/yr)} \\
\noalign{\smallskip}\hline\noalign{\smallskip}
Cha\,I    & 11 06 48 & $-$77 18 00 & 3.0 & $-24.3:-14.7$ & $-2.2:5.6$ \\  
Cha\,II   & 12 54 00 & $-$77 12 00 & 3.0 & $-33.2:-16.0$ & $-11.5:-1.9$    \\  
 % \noalign{\smallskip}
  \hline
\end{tabular}
\end{table*}
%$-$$-$$-$$-$$-$$-$$-$$-$$-$$-$$-$$-$$-$$-$$-$$-$$-$$-$$-$$-$$-$$-$$-$$-$$-$$-$$-$$-$$-$

%$-$$-$$-$$-$$-$$-$$-$$-$$-$$-$$-$$-$$-$$-$$-$$-$$-$$-$$-$$-$$-$$-$$-$$-$$-$$-$$-$$-$$-$
\addtocounter{table}{1} % For candidates
%$-$$-$$-$$-$$-$$-$$-$$-$$-$$-$$-$$-$$-$$-$$-$$-$$-$$-$$-$$-$$-$$-$$-$$-$$-$$-$$-$$-$$-$

%
%########################################################################
\section{New candidate members of the Cha~I and II moving groups}\label{sec:cand}

%_________________________________________________________________________
\subsection{Candidate selection}\label{sec:sel}

Using UCAC3 data, in Paper~I we identified four moving groups (i.e. groups of stars clustered around a given location in the proper motion diagram) 
toward the Chamaeleon sky area, corresponding to the stellar populations of the Cha~I and II clouds and of the foreground $\epsilon$ and $\eta$~Cha clusters (see Fig.~1 in that work). In this paper, we will focus only on the moving groups corresponding to the dark clouds (Cha~I and II), whose approximate locations are indicated by ellipses in Fig.~\ref{fig:cand}.

To perform a search of new candidate members of Cha~I and II, we first retrieved all objects in the UCAC4 catalogue within three degrees from the center of each cloud as given by SIMBAD (see Table~\ref{tab:search}). This radius value was chosen to maximize the search areas while avoiding spatial overlapping between them; in this way, we prevented the same source from being assigned to both clouds, as there is some overlapping in proper motion between the Cha~I and Cha~II moving groups.  Then, we defined two boxes in the vector point diagram covering each of the moving groups (see Fig.~\ref{fig:cand}); the box limits were set to 1$\sigma$ from the mean values obtained in Paper~I for each group. These selection criteria are summarized in Table~\ref{tab:search}. 

This step provided 3175 objects around Cha~I and  3831objects around Cha~II. Since the proper motion of a given star depends not only on its space motion tangential to the sky, but also on its distance, the retrieved lists include not only Chamaeleon members, but also foreground and background stars. Because the distance and radial velocities of virtually all sources are unknown, to purge the selection from contaminants we made use of the photometric information provided by UCAC4 (namely the Sloan $r'$ and $i'$ magnitudes from APASS, and the near-infrared $JHKs$ magnitudes from 2MASS), as this photometric set is available for all our objects.We constructed several colour-colour and colour-magnitude diagrams that have been shown to be useful tools to identify members of a given young association (see Fig.~\ref{fig:sel}). The location of the confirmed members in these diagrams served as a guide to identify new candidates. Only those objects whose location in all diagrams agreed with that of the confirmed members were selected. Objects brighter than $V\sim10.5$ were discarded, as they are probably saturated in APASS according to the catalogue description.\footnote{http://www.aavso.org/apass
} 
We note that no assumptions based on theoretical models were used for the selection. The APASS catalogue is complete down to $r\sim15.5$~mag and $i\sim15$~mag, which sets the completeness limit of this study. 

The final lists contained 51 objects in Cha~I and 14 in Cha~II, which are listed in Table~\ref{tab:cand}.\footnote{\footnotesize 
Obviously, the original samples also included most of the confirmed members, which were not included in this selection.}
Figure~\ref{fig:cand} shows their locations in the vector point diagram compared to those of the moving groups related to both clouds. Owing to the characteristics of the UCAC4 catalogue, we expect the selected sample to be biased towards low-extincted and relatively evolved sources,  because those objects in early stages of formation and evolution, or affected by high extinction, are very faint or undetectable in the optical. Furthermore, the lowest-mass component of both associations is missing, as these objects are too faint to be included in UCAC4. Objects with very strong H$\alpha$ emission may also be missed by our selection criteria, as such sources may appear bluer than expected for their age and spectral type in the diagrams of Fig.~\ref{fig:sel}. Examples of such objects can be seen among the confirmed members of Cha~II (right panels). 

The close distance and low Galactic latitude of the Chamaeleon clouds make us expect that contamination from both foreground and background objects is relatively low. However, because most of our studied area around each cloud is affected by virtually no extinction (see Sect.~\ref{sec:ext}), this contamination should be higher than observed within the cloud cores alone, where the cloud material itself blocks most of the background population.  The number of unrelated objects selected as candidate members in photometric surveys of nearby star-forming regions has been shown to be between 25\% and 75\% of the selection, depending on the location of the region within the Galaxy and the deepness of the survey. If proper motion information is available, such objects are almost completely eliminated, as shown in Paper~I and in our study of the Lupus clouds \citep{volupus, speclup}. 

In order to estimate the contamination within our samples, we performed several simulations of the Galaxy region in three degrees around the position of Cha~I and II using the Besan\c{c}on Galaxy Model 
\citep{robin1986,robin2003}. The simulations took the errors in photometry and proper motions into consideration. We repeated the selection process described in this section using the synthetic photometry and proper motions from the simulated catalogues, obtaining the number of objects expected to fulfill our criteria if no cluster were present in the direction of the Chamaeleon clouds. The resulting samples included between one and three objects in the Cha~I area and between zero and one in the Cha~II area, which would represent only up to  6\% and 7\%, respectively, of our candidate sample for each association. This result again suggests that the combination of astrometric and photometric information is very effective in reducing the contamination in searches for new members of star forming regions.

However, it must be noted that these simulations only took into account the astrometric errors derived from the photometric errors of the souces. Source confusion or centroiding problems may lead to incorrect estimations of the proper motions, effectively increasing the amount of contamination in our samples. Although the UCAC4 flags allow us to get rid of most of the sources with uncertain or incorrect proper motions, some of these objects may still be included in our selected samples. Unfortunately, it is not easy to estimate the amount of this contamination. Because these problems are most likely to happen in the bright and faint end of the UCAC4 magnitude range, we expect it to be relatively high, given the number of objects with $r'$ magnitudes larger than 14. On the other hand, we may also miss members of the Cha~I and II moving groups owing to incorrectly estimated proper motions in UCAC4 that place them outside the boundaries of our selection boxes in Fig.~\ref{fig:sel}.  

%$-$$-$$-$$-$$-$$-$$-$$-$$-$$-$$-$$-$$-$$-$$-$$-$$-$$-$$-$$-$$-$$-$$-$$-$$-$$-$$-$$-$$-$
\addtocounter{table}{1} % For Simbad objects
%$-$$-$$-$$-$$-$$-$$-$$-$$-$$-$$-$$-$$-$$-$$-$$-$$-$$-$$-$$-$$-$$-$$-$$-$$-$$-$$-$$-$$-$

%_________________________________________________________________________
\subsection{SIMBAD objects}\label{sec:simbad}

To look for additional information about our selected objects, we checked whether any of our new candidates were known to SIMBAD. 

This was the case for  seventeen stars from the Cha~I list and four from the Cha~II list, representing neartly one third (about 33\% and 28.5\%, respectively) of the total samples. These objects are listed in Table~3, %\ref{tab:simbad}. 
along with their reported proper motions, radial velocities and spectral types, if available.

Nearly all the stars in Table~3 %\ref{tab:simbad}
have reported proper motions in the Tycho-2 catalogue \citep{hog2000},
which are statistically compatible with the UCAC4 measurements within 3$\sigma$. Even so, the Tycho-2 proper motions place all but three objects outside the corresponding selection box defined in Table~\ref{tab:search}. The exceptions are \object{ChaI-PM-5} (\object{TYC 9418-388-1}), \object{ChaI-PM-38} (\object{2MASS~J11475667-7536316}), and \object{ChaII-PM-12} (\object{TYC~9413-205-1}), for which the Tycho-2 proper motions are still within the selection box in the vector-point diagram. It must be noted, however, that the Tycho-2 measurements are less precise than those of UCAC4 for most of the stars listed in Table~3, %\ref{tab:simbad} 
owing to the lower number of epochs used to compute the proper motion, and because nearly all the sources are fainter than the completeness limit of the Tycho-2 survey ($V\sim11$~mag). Therefore, we decided to keep these objects in the subsequent analysis.

For one star, namely \object{ChaI-PM-13} (\object{2MASS~J110593816-782421}), there is also a radial velocity measurement available from the RAVE survey \citep{siebert2011}, which is slightly lower than the mean for Cha~I stars reported in Paper~I, a further suggestion that this is likely to be a contamination object.

Three SIMBAD sources, namely \object{2MASS~J11120518-7712408} (\object{ChaI-PM-20}), \object{IRAS 12348-7634} (\object{ChaII-PM-9}), and \object{CR~Mus} (\object{ChaII-PM-18}) do not have reported Tycho-2 proper motions or RAVE radial velocities. The third object  is a known Mira variable, and hence a sure contaminant.

%
%########################################################################
\section{SED analysis}\label{sec:vosa}

To further purge our candidate lists of likely contaminants, and to learn more about the nature of our objects, we compiled all the photometry available for them in the Virtual Observatory and in other public archives. With these data, we constructed the spectral energy distributions (SEDs), which were analysed to derive their fundamental parameters using the latest version of the VO SED Analyzer tool\footnote{\footnotesize http://svo2.cab.inta-csic.es/svo/theory/vosa/}
\citep[VOSA;][]{bayo2008, bayo2012}.

%_________________________________________________________________________
\subsection{Photometry compilation and observed SEDs}\label{sec:phot}

%$-$$-$$-$$-$$-$$-$$-$$-$$-$$-$$-$$-$$-$$-$$-$$-$$-$$-$$-$$-$$-$$-$$-$$-$$-$$-$$-$$-$$-$
\addtocounter{table}{1} % For photometry
%$-$$-$$-$$-$$-$$-$$-$$-$$-$$-$$-$$-$$-$$-$$-$$-$$-$$-$$-$$-$$-$$-$$-$$-$$-$$-$$-$$-$$-$

Most of the multiwavelength photometry was retrieved with VOSA from a number of public databases available through the Virtual Observatory. These data include optical photometry from Tycho-2 \citep[$B$ and $V$ bands;][]{hog2000}  and DENIS \citep[$I$ band; ][]{denis2005}, near-infrared $JHKs$ photometry from 2MASS \citep{skrutskie2006}, and mid-infrared photometry from AKARI and WISE \citep{ishihara2010, cutri2012}.  For the Cha~II candidates, we also retrieved the IRAC and MIPS photometry (from 3.6 to 24~$\mu$m) from the {\em ``Cores to Disks''} ({\em c2d}) {\em Spitzer} legacy program \citep{alcala2008}, also available through VOSA. 

Additional {\em Spitzer}/IRAC  photometry for a few Cha~I candidates was obtained after retrieving and processing all available observations from the IRSA database (programs 6, 36, 37, 139, 173, and 30574). We retrieved the corresponding BCD images and associated ancillary products from the public archive and processed them following standard procedures with the recommended MOPEX software \citep{makovoz2005}. The IRAC observations were obtained with various exposure times (0.6~s, 12~s, 30~s, and 100~s) and the final sensitivity depends on the coverage and observation history. We processed each set independently to cover the widest dynamic range. The procedure within MOPEX includes overlap correction, resampling, interpolation (to have an output pixel scale of 0\farcs6), and outlier rejection. The individual frames were then median-combined with {\it Swarp} \citep{bertin2002} using the rms maps produced by MOPEX as weight maps. The resulting photometric catalogues were cross-matched with our list of Cha~I kinematical candidate members.

The compiled multiwavelength photometry (also including the BV$g'r'i'$ APASS photometry from UCAC4) for all the objects is summarized in Table~\ref{tab:phot}.\footnote{Available online.} The data were used to construct the observed SED for each object, which was then visually inspected to check the consistency of the photometric datapoints and to examine its shape. Most of the candidates display curves suggesting they are bare (diskless) photospheres. This impression is confirmed by VOSA, as the tool does not detect any infrared excess for most of the objects. Some sources (e.g. \object{ChaI-PM-48}) could have a little excess at the longest wavelengths ($\gtrsim8~\mu$m), usually associated with the presence of evolved disks. We discuss these objects later in Sect.~\ref{sec:fits}.

%_________________________________________________________________________
\subsection{Extinction maps}\label{sec:ext}

%$-$$-$$-$$-$$-$$-$$-$$-$$-$$-$$-$$-$$-$$-$$-$$-$$-$$-$$-$$-$$-$$-$$-$$-$$-$$-$$-$$-$$-$
   \begin{figure*}[ht]
   \centering
     \includegraphics[width=\textwidth]{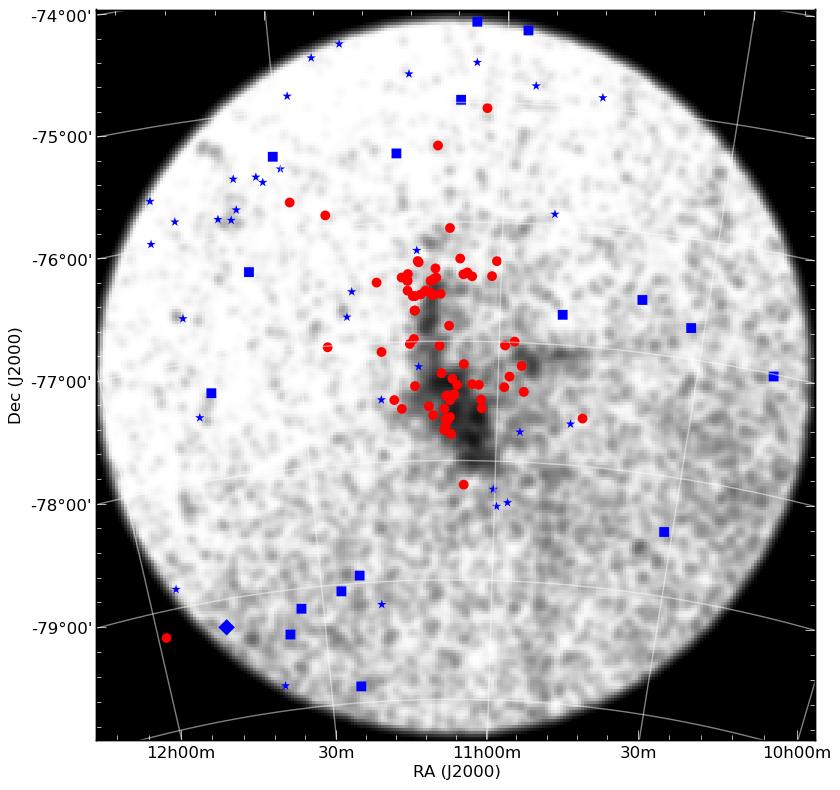}\hfill
     \caption{\footnotesize
	       Extinction map of the Cha~I cloud created with the adaptive grid method described in Sect.~\ref{sec:ext}. The kinematic members identified in Paper~I are overplotted as red circles. The objects finally selected as probable and possible members of the moving groups  after the SED analysis are marked as blue squares and diamonds, respectively, and the rest of kinematical candidates selected on the basis of proper motions and colours are indicated as blue stars (see Sect.~\ref{sec:lists} for details).
	       }
         \label{fig:chaimap}
   \end{figure*}
%$-$$-$$-$$-$$-$$-$$-$$-$$-$$-$$-$$-$$-$$-$$-$$-$$-$$-$$-$$-$$-$$-$$-$$-$$-$$-$$-$$-$$-$

%$-$$-$$-$$-$$-$$-$$-$$-$$-$$-$$-$$-$$-$$-$$-$$-$$-$$-$$-$$-$$-$$-$$-$$-$$-$$-$$-$$-$$-$
   \begin{figure*}[ht]
   \centering
     \includegraphics[width=\textwidth]{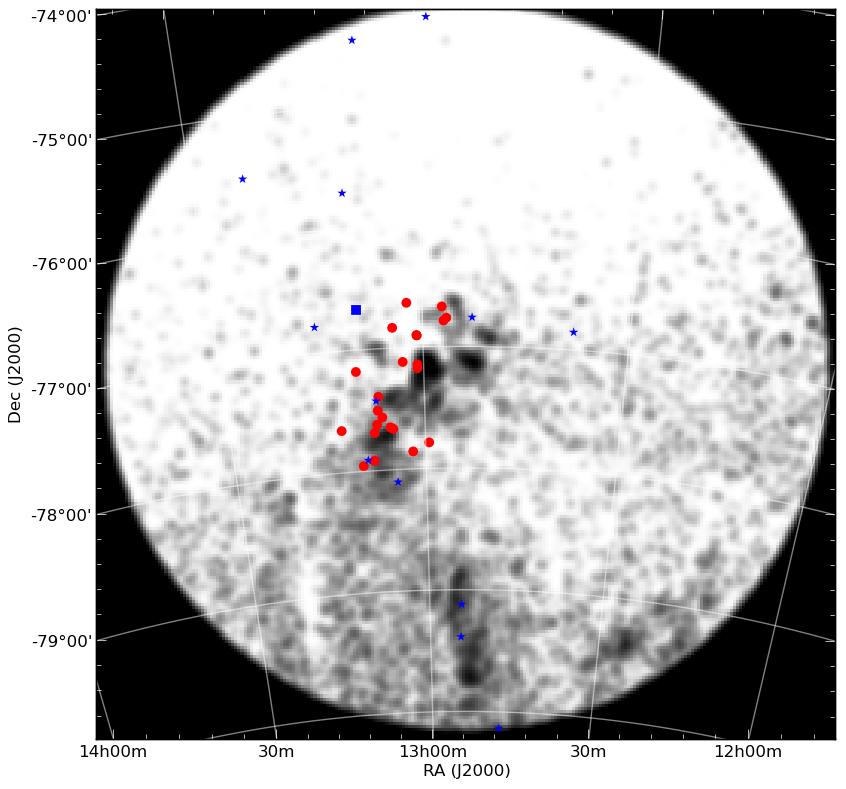}
     \caption{\footnotesize
	       Same as Fig.~\ref{fig:chaimap} for the Cha~II cloud.
	       }
         \label{fig:chaiimap}
   \end{figure*}
%$-$$-$$-$$-$$-$$-$$-$$-$$-$$-$$-$$-$$-$$-$$-$$-$$-$$-$$-$$-$$-$$-$$-$$-$$-$$-$$-$$-$$-$

We followed the adaptive grid method developed by \citet{cambresy1997} to derive extinction maps for the sky areas around the Cha~I and II clouds where our new candidates are located. This method is based on the voids induced by star forming regions in the uniform distribution of background stars, clearly observable with all-sky surveys. We used 2MASS as the parent catalogue, and we had a control field at a similar Galactic latitude to model the background population. The average pixel size of the maps is 1.5$^{\prime}$, although it varies depending on the extinction itself, as the grid of coordinates is built according to the local density of sources. Comparison with maps of the same regions by other authors \citep{cambresy1997,kainulainen2009} shows very good agreement for areas with $A_V\le12$~mag. As mentioned earlier, UCAC4 is based on optical catalogues, and so we are not likely to find candidates in these very heavily extincted areas. Further details on the adaptive grid method are given in \citet{bayo2012}. 

We show our extinction maps in Figs.~\ref{fig:chaimap} and \ref{fig:chaiimap}, where we have also plotted the projected spatial locations of the confirmed members and of our new candidates. Because our candidates are mostly located outside the dark clouds, extinction towards them is expected to very low. We used the information from the maps to derive extinction estimates ($A_V^{grid}$) along the line of sight towards the location of each of our candidates, which are listed in Table~\ref{tab:cand}. The extinction values have an estimated error of 0.05~mag, derived from variations in the control field; therefore, when a zero value is indicated, it is implied that  $A_V^{grid}\lesssim$0.05~mag.

With very few exceptions, our candidates are not affected by extinction according to our extinction maps. Only a few objects are located in areas of low-to-moderate extinction, the highest value corresponding to \object{ChaII-PM-1} ($A_V^{grid}=2.81$~mag).

It is important to note that the adaptive grid method provides an estimate of the total extinction in a given direction. Thus, in an area of high extinction such as the dark cloud cores, $A_V^{grid}$ can be assumed to be the true $A_V$ of the star only if it is located behind the cloud \citep[for a detailed discussion, see][]{bayo2012}. If it is placed before or within the cloud, the amount of dust will be lower and the star will be less extincted than suggested by the maps. On the other hand, some intrinsic extinction may be present in young stellar objects surrounded by large amounts of circumstellar material, but this does not seem to be the case for the vast majority of our candidates, as they are diskless according to their SEDs (see Sect.~\ref{sec:phot} above). Thus,  without information on the individual distance, the $A_V^{grid}$ value should be regarded as an upper limit to the real extinction affecting a particular object, and the true $A_V$ may in principle have any value between 0 and $A_V^{grid}$.

%_________________________________________________________________________
\subsection{SED fitting}\label{sec:fits}

%----------------------------------------------------------
\begin{table}[t]
\label{tab:fitpars}
\caption{Input parameter spaces for the SED fitting}
\centering
\begin{tabular}{l c c}
\hline
\hline\noalign{\smallskip}
  \multicolumn{1}{l}{Parameter} &
  \multicolumn{2}{c}{Models$^{\mathrm{a}}$} \\
\noalign{\smallskip}
  \multicolumn{1}{l}{} &
  \multicolumn{1}{c}{BT-Settl} &
  \multicolumn{1}{c}{Kurucz}  \\
\noalign{\smallskip}\hline\noalign{\smallskip}
 $T_{eff}$ (K) $^{\mathrm{b}}$    & $T_{min}$-4500 & 4500-50000  \\  			   
 $\log g$             & 3.5-5.0          & 3.5-5.0  \\  			   
 Metallicity          & 0.0                 & 0.0    \\  			   
 $A_V$ (mag) $^{\mathrm{c}}$    & 0.0-$A_V^{grid}$        & 0.0-$A_V^{grid}$    \\  			   
  \noalign{\smallskip}\hline\noalign{\smallskip}
\end{tabular}
\begin{flushleft}
{\bf Notes.} \\
     $^{\mathrm{a}}$ References: \citet{allard2012, castelli1997} \\
     $^{\mathrm{b}}$ $T_{min}=$2700~K for Cha~I and 2800~K for Cha~II (see text). \\
     $^{\mathrm{c}}$ Individual $A_V^{grid}$  values given in Table~\ref{tab:phot}. \\  
\end{flushleft}
\end{table}
%----------------------------------------------------------

%$-$$-$$-$$-$$-$$-$$-$$-$$-$$-$$-$$-$$-$$-$$-$$-$$-$$-$$-$$-$$-$$-$$-$$-$$-$$-$$-$$-$$-$
\addtocounter{table}{1} % For fitting results and physical parameters
%$-$$-$$-$$-$$-$$-$$-$$-$$-$$-$$-$$-$$-$$-$$-$$-$$-$$-$$-$$-$$-$$-$$-$$-$$-$$-$$-$$-$$-$

The SEDs were fitted to theoretical models in order to estimate effective temperatures, luminosities, and surface gravities for the candidates, testing the parameter space summarized in Table~5. %\ref{tab:fitpars}. 
The input for the fits included the object name, the coordinates, the user-provided and VO photometry, and a distance value. 

Since we do not have parallax measurements for our objects, the distance was set to the reported mean distances to the clouds: $160\pm15$~pc in the case of Cha~I and $178\pm18$~pc for Cha~II \citep{whittet1997}. In addition, the information from the extinction maps was used to set the ranges of visual extinction ($A_V$) to be tested by the procedure. Because most of our objects are seen towards areas of nearly negligible extinction ($\lesssim$0.05~mag), in practice we set the extinction $A_V=0$ for the majority of them. For those objects with $A_V^{grid}>0.05$~mag, we tested the interval between 0 and the individual $A_V^{grid}$ values in steps of $A_V^{grid}/20$, the VOSA default step when the extinction is not set to a fixed value by the user.

The BT-Settl models \citep{allard2012} were used to test the range of effective temperatures between 4500~K and the minimum effective temperature expected for our candidates, should they be young stars located at the distance of the dark clouds. This minimum temperature corresponds to a very low-mass star with I-band magnitude in the lower limit of our candidate lists ($\sim$16.5~mag) at the distance of each cloud, and was estimated to be about 2700~K and 2800~K for Cha~I and II, respectively, using the \citet{baraffe1998} evolutionary tracks. We assumed in this estimation that the objects are not older than 50~Myr, and that they are not extincted. The temperature of 4500~K, on the other hand, is the predicted temperature for TiO formation in the atmospheres of low-mass stars. \citet{allard2012} claim that the BT-Settl models have a better treatment of the oxygen abundance than any previous models and, therefore, they provide better approximations below 4500~K. The Kurucz models \citep{castelli1997} were used to test the range between this temperature and 50000~K, the maximum value provided by this grid. For the few objects having best-fitting temperatures of precisely 4500~K, we checked that the results using both models were consistent; because the Kurucz models usually provided better fits (according to the $\dot{\chi}^2$ statistics, see below), the best-fitting model from this grid was kept as best fit in those cases. 

A range of gravities was tested to account for the possibility that the stars have a variety of ages and evolutionary stages; however, the results depended very weakly on the gravity values. Only solar metallicity was considered, in agreement with the results by \citet{santos2008} for some previously known Cha~I members.

Because most of the stars do not have infrared excesses, the number of datapoints to fit was relatively high (typically 12 to 15). If an excess was detected, the affected datapoints were not considered in the fitting process. Because the possible excesses all started at relatively long wavelengths, the number of datapoints left was always enough to perform the fit.

The goodness of the fits was measured by two different statistics: reduced $\chi^2$ minimization ($\dot{\chi}^2$)  and Bayesian probability analysis of the model parameters \citep[for details, see][]{bayo2008, bayo2012}. Because several models provided good approximations to a given SED with only very small differences in the $\dot{\chi}^2$ values, the information from the Bayesian analysis helped us to choose the most appropriate model by selecting the one whose parameter values had the greatest probability. This model was usually (but not always) the one with the lowest $\dot{\chi}^2$. 

For a few objects, the $T_{eff}$ value with maximum probability is namely in the boundary of the considered range, with the probability distribution monotonically increasing toward this value. This may happen if the actual most probable temperature value is out of the tested range (such as a $T_{eff}$ lower than 4500~K in the case of the Kurucz models), because the closest value  is chosen by the algorithm in this case. Therefore, the results of the Bayesian analysis are not trustworthy in these cases, and we had to rely only on the $\dot{\chi}^2$ values to decide on the goodness of the fits. 

The process provides very good approximations to the SEDs, with $\dot{\chi}^2\lesssim10$, for 35\% of the Cha~I sample and 14\% of the Cha~II sample (18 and 2 objects, respectively). We checked that the best-fitting values were the ones with highest Bayesian probabilities, and that the $T_{eff}$ had probability $>0.5$ \citep[see also discussion in][]{bayo2012}. The fits that did not fulfill these criteria were considered bad, and the objects were not further considered.

The best-fitting results for our candidates are summarized in Table~\ref{tab:fitres}, where both the value of the $\dot{\chi}^2$ and the Bayesian probabilities are indicated.

%$-$$-$$-$$-$$-$$-$$-$$-$$-$$-$$-$$-$$-$$-$$-$$-$$-$$-$$-$$-$$-$$-$$-$$-$$-$$-$$-$$-$$-$
   \begin{figure}[t]
   \centering
       \includegraphics[width=0.55\textwidth]{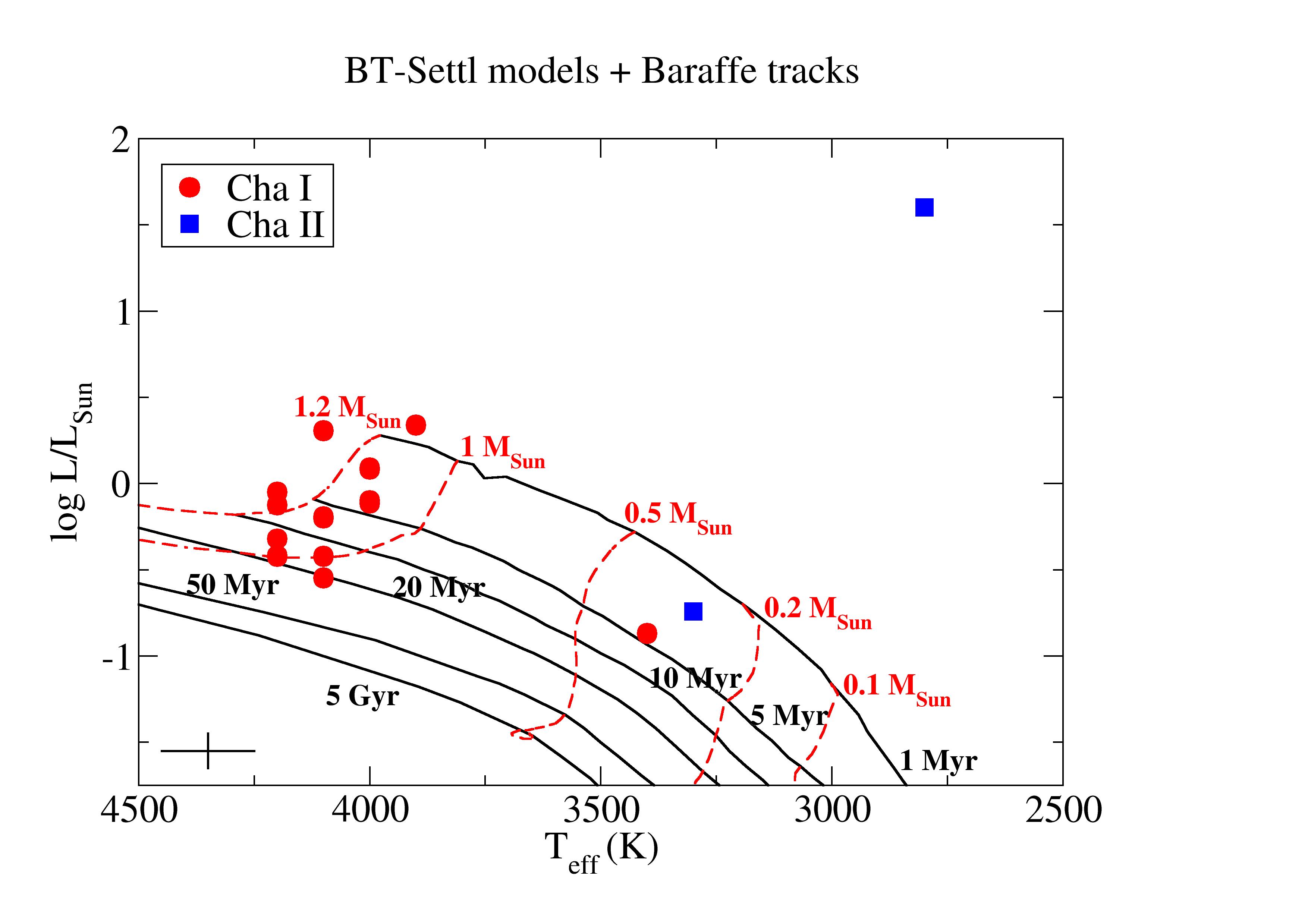}\hfill
\caption{\footnotesize
		     Hertzprung-Russell diagram for our candidate members of Cha~I and II with good fits, compared to the isochrones and mass tracks from \citet{baraffe1998} (solid black lines and dashed red lines, respectively). The plotted isochrones correspond to ages of 1, 5, 10, 20, 50~Myr and 5~Gyr (from top to bottom), and the mass tracks  to 1.2, 1, 0.5, 0.2, and 0.1$M_{\odot}$ (from left to right). The cross in the bottom-left corner of the diagram indicates the typical error.			    
		      }
         \label{fig:hr}
   \end{figure}
%$-$$-$$-$$-$$-$$-$$-$$-$$-$$-$$-$$-$$-$$-$$-$$-$$-$$-$$-$$-$$-$$-$$-$$-$$-$$-$$-$$-$$-$

%----------------------------------------------------------
\setcounter{table}{6}
\begin{table*}[t]
\label{tab:means}
\caption{UCAC4 mean and weighted mean proper motions for the confirmed and probable members of the Cha~I and II moving groups}
\centering
\begin{tabular}{l l l l l l l }
\hline
\hline\noalign{\smallskip}
  \multicolumn{1}{l}{Group} &
  \multicolumn{3}{c}{Arithmetic means} &
  \multicolumn{3}{c}{Weighted means} \\
  \multicolumn{1}{l}{} &
  \multicolumn{1}{c}{$<\mu_{\alpha}\cos\delta>$} &
  \multicolumn{1}{c}{$<\mu_{\delta}>$} &
  \multicolumn{1}{c}{$<\mu>$} &  
  \multicolumn{1}{c}{$<\mu_{\alpha}\cos\delta>$} &
  \multicolumn{1}{c}{$<\mu_{\delta}>$} &
  \multicolumn{1}{c}{$<\mu>$}  \\
  \multicolumn{1}{l}{} &
 \multicolumn{1}{c}{(mas/yr)} &
 \multicolumn{1}{c}{(mas/yr)} &
  \multicolumn{1}{c}{(mas/yr)} &
 \multicolumn{1}{c}{(mas/yr)} &
 \multicolumn{1}{c}{(mas/yr)} &
  \multicolumn{1}{c}{(mas/yr)} \\
\noalign{\smallskip}\hline\noalign{\smallskip}
  \noalign{\smallskip}
 \object{Cha~I confirmed members}  & $-$19.2\,$\pm$\,1.8 &   \hspace{0.15cm}+1.5\,$\pm$\,2.1 & 19.4\,$\pm$\,1.7 & $-$19.2\,$\pm$\,0.2 &   \hspace{0.15cm}+1.7\,$\pm$\,0.3 & 19.60\,$\pm$\,0.15 \\ 
 \object{Cha~I probable members}   & $-$18.8\,$\pm$\,2.8 &   \hspace{0.15cm}+2.3\,$\pm$\,2.0 & 19.0\,$\pm$\,2.9 & $-$19.0\,$\pm$\,0.3 &   \hspace{0.15cm}+2.5\,$\pm$\,0.3 &  19.5\,$\pm$\,0.3  \\ 
 \object{Cha~II confirmed members} & $-$21.6\,$\pm$\,3.7 & \hspace{0.15cm}$-$7.0\,$\pm$\,2.8 & 22.9\,$\pm$\,3.6 & $-$21.0\,$\pm$\,0.7 & \hspace{0.15cm}$-$7.1\,$\pm$\,0.8 &  22.4\,$\pm$\,0.9  \\ 
 \object{Cha~II probable member$^{\mathrm{a}}$}   & $-$23.0\,$\pm$\,4.6 & \hspace{0.15cm}$-$8.2\,$\pm$\,4.5 & 24.4\,$\pm$\,5.8 & $-$23.0\,$\pm$\,4.6 & \hspace{0.15cm}$-$8.2\,$\pm$\,4.5 &  24.4\,$\pm$\,5.8  \\ 
\noalign{\smallskip}\hline\noalign{\smallskip}
\end{tabular}
\begin{flushleft}
{\bf Notes.} \\
    $^{\mathrm{a}}$  The individual measurements are given in this case.  \\
\end{flushleft}
\end{table*}
%----------------------------------------------------------

%_________________________________________________________________________
\subsection{Mass and age estimations}\label{sec:hr}

%$-$$-$$-$$-$$-$$-$$-$$-$$-$$-$$-$$-$$-$$-$$-$$-$$-$$-$$-$$-$$-$$-$$-$$-$$-$$-$$-$$-$$-$
\addtocounter{table}{1} % For parameters
%$-$$-$$-$$-$$-$$-$$-$$-$$-$$-$$-$$-$$-$$-$$-$$-$$-$$-$$-$$-$$-$$-$$-$$-$$-$$-$$-$$-$$-$

The best-fitting effective temperatures and bolometric luminosities allowed us to place the objects with good fits in the Hertzprung-Russell diagram (Fig.~\ref{fig:hr}). This way, it was possible to derive masses and ages for those candidates and to identify the likely members of the Cha~I and II clouds using theoretical evolutionary tracks. We used the \citet{baraffe1998} set of isochrones for this analysis. The luminosities, masses, and ages derived in this way are summarized in Tables~\ref{tab:fitres}. According to our estimations, the objects for which this exercise was possible are not older than 20~Myr and have masses between 0.3 and 1.3$M_{\odot}$, supporting that they are likely young, low-mass stars. Given the estimated errors in $\log L$ and $T_{eff}$ (about 0.1 dex and 100~K in average, respectively), we estimate that the errors in age are about 1, 2 and 5~Myr for objects younger than 5, 10 and 20~Myr, respectively, and that the errors in mass are of about 0.1$M_{\odot}$. We caution, however, that the errors in $\log L$ and $T_{eff}$ do not take stellar variability or multiplicity into account, so the actual errors may be higher, and thus they may be largely responsible for the age scatter we obtain. The effect of multiplicity and variability in the VOSA results are discussed in detail in  \citet{bayo2011} and \citet{bayo2012}, respectively.

For two of our candidates, namely \object{ChaI-PM-4} and \object{ChaII-PM-14}, it was not possible to derive mass and age values, as the SED fitting results place them above the 1~Myr isochrone in the Hertzprung-Russell diagram. While this location is usually interpreted as an indication of a very young age, we think it unlikely in these cases, as neither of the objects possesses a thick disk according to the SED analysis. In the case of \object{ChaI-PM-4}, given its location close to the 1~Myr isochrone in Fig.~\ref{fig:hr}, the errors in $T_{eff}$ and $\log L$ still leave open the possibility that it is a young member of the Cha~I moving group.  We note, however, that this object has reported Tycho-2 proper motions completely discrepant from the UCAC4 values, which would be in disagreement with membership to the Cha~I moving group.

Object \object{ChaII-PM-14}, on the other hand, is seen so much above the 1~Myr isochrone that it looks much more likely to be a misplacement owing to an incorrect distance estimate. In addition, its best-fitting effective temperature is 2800~K, namely on the lowest boundary of the interval tested with VOSA for the Cha~II candidates, with an estimated Bayesian probability of 100\%. As discussed in Sect.~\ref{sec:fits}, this may be an indication that the true effective temperature of this object is actually beyond this boundary value, which is in complete disagreement with  \object{ChaII-PM-14} belonging to the Cha~II moving group. Taken all together, these results suggest a low probability of  \object{ChaII-PM-14} being a young star.

%
%########################################################################
\section{Final candidate lists}\label{sec:lists}

After the results of the SED analysis described above, we classify as probable members of the moving groups (labelled ``m") in Table~\ref{tab:fitres}) those stars with reliable fits and derived ages younger than about 100~Myr. This way, from 51 objects reported in Sect.~\ref{sec:sel}, we identify 17 probable members of the Cha~I moving group. Following the discussion from the previous Section, \object{ChaI-PM-4} is considered a possible member (label  ``m?"). The rest of objects from the Cha~I candidate sample remain unclassified owing to unreliable fits and insufficient or contradictory additional information (label  ``nc").

In the Cha~II sample, only one star (\object{ChaII-PM-11}) out of the 14 objects selected in Sect.~\ref{sec:sel} is classified as a probable member. The other source with a good fit, \object{ChaII-PM-14}, is rejected as a probable non-member (label  ``nm"). Following the discussion from Sect.~\ref{sec:simbad}, also \object{ChaII-PM-18} is rejected as a probable non-member. The remaining 11 sources are not classified.

With these figures, the number of identified contaminants (non-members) is null in the Cha~I sample and amounts to about 14\% of the Cha~II sample. The former figure is lower, and the latter figure is higher than expected from the simulations described in Sect.~\ref{sec:sel}. However, because more than 60\% of our candidates could not be classified with the SED analysis, the true contamination may be higher in both clouds.

The final (probable and possible) candidate members are marked with solid symbols in the vector-point diagram of Fig.~\ref{fig:cand}. In  Table~7, %\ref{tab:means}, 
we show the arithmetic and weighted mean proper motions of the probable member samples. The table also lists the corresponding means for the confirmed members of both moving groups; they have been recalculated using UCAC4 data, but are compatible (although slightly higher) with the values given in Paper~I.

%
%########################################################################
\section{Properties of the new candidate members}\label{sec:prop}
 
%$-$$-$$-$$-$$-$$-$$-$$-$$-$$-$$-$$-$$-$$-$$-$$-$$-$$-$$-$$-$$-$$-$$-$$-$$-$$-$$-$$-$$-$
\addtocounter{table}{1} % For parameters of members
%$-$$-$$-$$-$$-$$-$$-$$-$$-$$-$$-$$-$$-$$-$$-$$-$$-$$-$$-$$-$$-$$-$$-$$-$$-$$-$$-$$-$$-$

We now discuss the properies of our new candidate members of Cha~I and II, and compare them to the spectroscopic members confirmed in Paper~I  (73 and 22 objects from the Cha~I and II moving groups, respectively). We remark that the member samples do not include all the known objects in the dark clouds, but only those optically detected and with available proper motions in UCAC; hence, they are expected to be biased towards relatively evolved and massive objects, and towards low extinctions.

%_________________________________________________________________________
\subsection{Spatial position}\label{sec:dist}

Figure~\ref{fig:chaimap} shows the projected location on our extinction map of the confirmed members and our new candidate members to the \object{Cha~I} moving group. While the confirmed members tend to cluster within the cloud cores, most of the Cha~I candidates are seen spread around the dark cloud, mostly to the north and east. This distribution may be partially caused by the fact that the south and west of our surveyed area are more extincted than the north and east. Even so, it is interesting to note that this is the direction towards the Galactic plane and towards the \object{Musca} filament and the \object{Southern Coalsack}, two regions to which the \object{Chamaeleon} cloud complex has been related \citep{king1979}.

Most of our new candidates are sharing locations with known members at the outskirts of the cloud, or seem to bridge the gap between the cloud population and a few known dispersed members to the north and south-east. Most of our new probable members (that is, those objects with fits considered reliable and derived physical parameters consistent with young stars) are located in these two groupings. The south-eastern group is particularly interesting, as it is located halfway to the dark cloudlet \object{[DB2002b]\,G300.23-16.89}, in an area where many of the $\epsilon$~Chamaleontis cluster members are found. Interestingly, the only confirmed Cha~I member in the south-east group, \object{RXJ1158.5-7913}, is seen projected toward this small cloudlet, and detached from any other confirmed Cha~I members; nevertheless, the object was shown in Paper~I to have kinematic properties that are in much better agreement with membership to  the Cha~I moving group than to the $\epsilon$~Cha group. This membership would be reinforced if our new candidates are confirmed as young low-mass stars. This would in turn  strengthen the idea that the  [DB2002b]\,G300.23-16.89 cloud is related to the Chamaeleon complex, or at least to the Cha~I cloud, as previously suggested in the literature \citep{whittet1997}.

The spatial distribution of the \object{Cha~II} members and candidate members is shown in Fig.~\ref{fig:chaiimap}. Most of our new candidates are located to the north-east of the dark cloud. Again, as in the case of Cha~I, this is the direction towards the Galactic plane and the Musca and Southern Coalsack regions, but also the less-extincted area in the surveyed region around the dark cloud. Only a few objects (\object{ChaII-PM-6}, \object{ChaII-PM-7}, \object{ChaII-PM-8}, and to some extent also \object{ChaII-PM-10}) are located in the same areas as the confirmed members of the moving group. The only probable member in the sample is also seen not far from the confirmed members, to the north-east of the northern dark cloud core. 

Interestingly, three Cha~II candidates, namely  \object{ChaII-PM-1}, \object{ChaII-PM-2}, and \object{ChaII-PM-3}, are seen to the south of the Cha~II dark cloud, in or near the dark cloud \object{[DB2002b]\,G303.11-16.08} that seems to bridge Cha~II and III. \object{Cha~III} itself lies outside our search area.

%$-$$-$$-$$-$$-$$-$$-$$-$$-$$-$$-$$-$$-$$-$$-$$-$$-$$-$$-$$-$$-$$-$$-$$-$$-$$-$$-$$-$$-$
   \begin{figure}[t]
   \centering
     \includegraphics[width=0.5\textwidth]{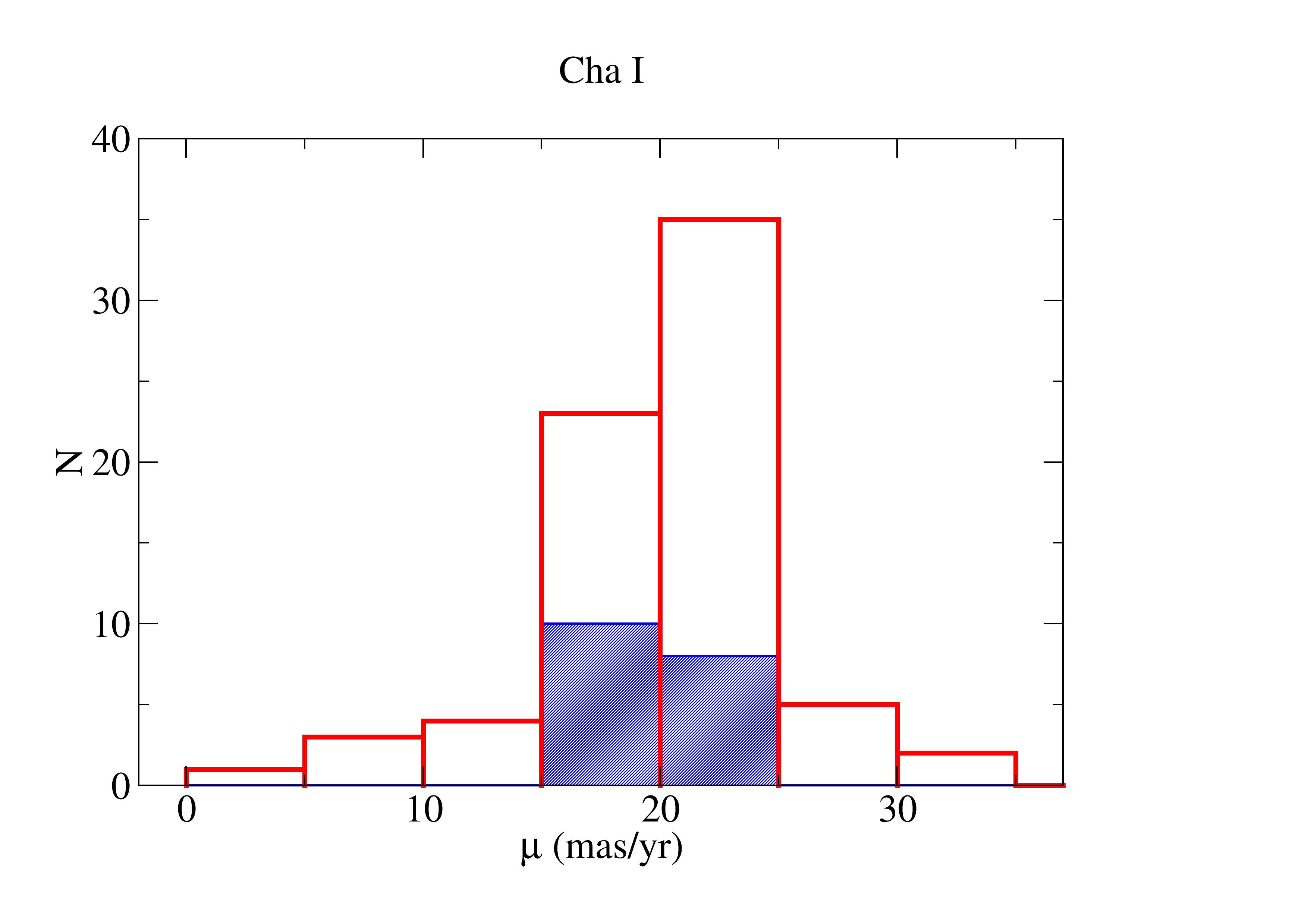}\hfill
     \caption{\footnotesize
	      Total proper motion distribution of the new candidate members (blue hashed histogram) and the confirmed members from Paper~I (red line histogram) for the Cha~I moving group.	      
	       }
         \label{fig:pm}
   \end{figure}
%$-$$-$$-$$-$$-$$-$$-$$-$$-$$-$$-$$-$$-$$-$$-$$-$$-$$-$$-$$-$$-$$-$$-$$-$$-$$-$$-$$-$$-$

%_________________________________________________________________________
\subsection{Proper motions}\label{sec:pm}

Figure~\ref{fig:cand} shows the locations of our new candidate members to the Cha~I and Cha~II moving groups in the vector-point diagram. In the Cha~I group, the objects tend to cluster in the right half of the selection box depicted in that diagram. However, when only the probable members (objects with good fits) are considered, this trend is no longer evident. This is confirmed by the left panel of Fig.~\ref{fig:pm}, which shows the proper motion histograms for the confirmed and probable members. As shown in Table~7, the mean values of the candidate sample (($-18.8\pm2.8, 2.3\pm2.0$)~mas/yr) are fully consistent with the means for the confirmed Cha~I members.

Also our new candidate members to the Cha~II moving group tend to be located to the right half of the selection box in the vector point diagram, while one third of the box to the left is devoid of objects. The only probable member in the Cha~II sample is located in the lower central part of the box. The values listed in Table~7 confirm that this source has proper motion values similar to the means for the kinematical cloud members.

%$-$$-$$-$$-$$-$$-$$-$$-$$-$$-$$-$$-$$-$$-$$-$$-$$-$$-$$-$$-$$-$$-$$-$$-$$-$$-$$-$$-$$-$
   \begin{figure*}[t]
   \centering
     \includegraphics[width=\textwidth]{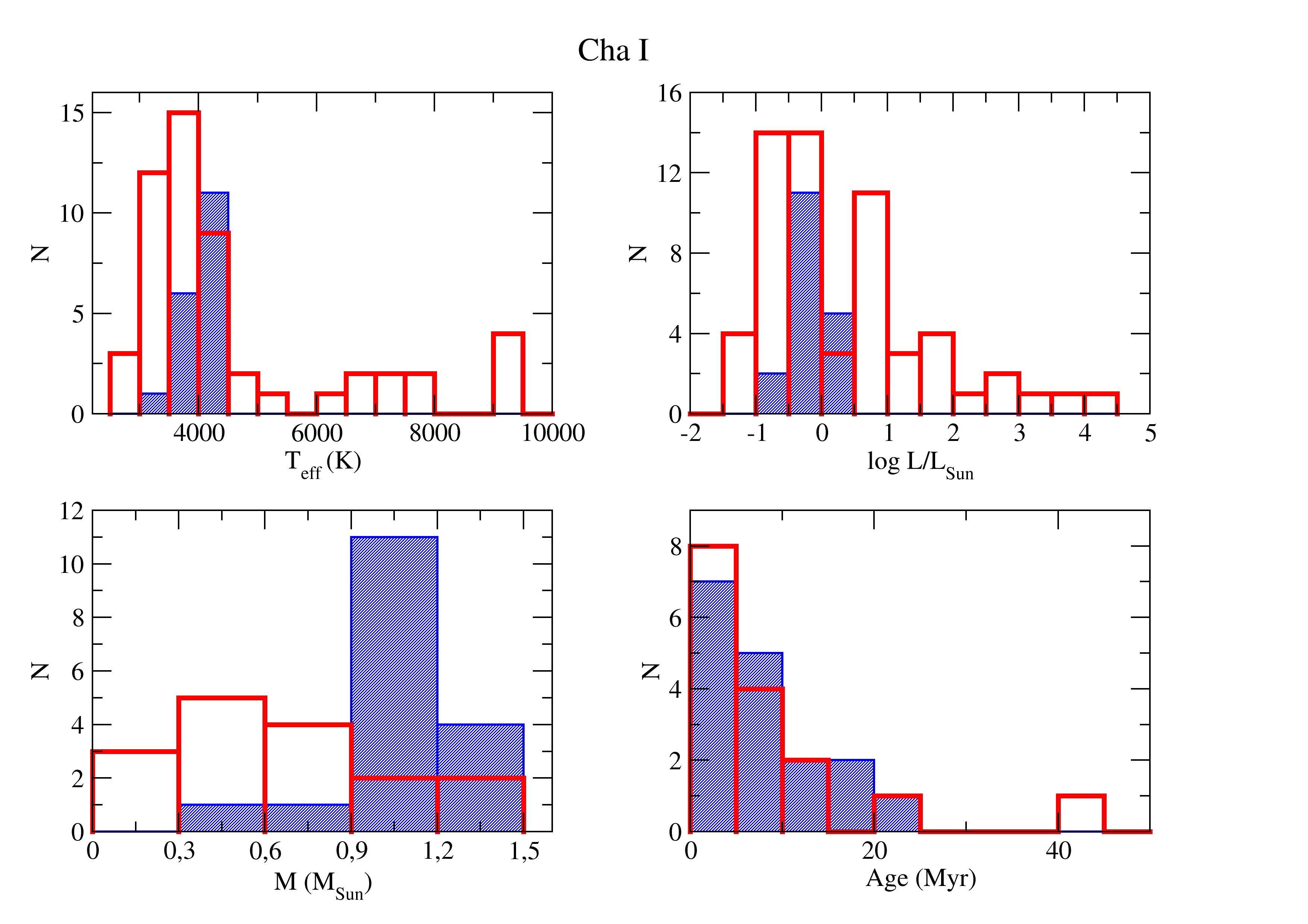}\hfill
     \caption{\footnotesize
	       Distributions of the physical parameters obtained from the SED fitting for our new candidate members of the Cha~I moving group (blue hashed histogram), compared to that of the confirmed members from Paper~I (red line histogram). We note that the samples are biased towards relatively evolved, low-extincted objects.
	       }
         \label{fig:chaipars}
   \end{figure*}
%$-$$-$$-$$-$$-$$-$$-$$-$$-$$-$$-$$-$$-$$-$$-$$-$$-$$-$$-$$-$$-$$-$$-$$-$$-$$-$$-$$-$$-$

%_________________________________________________________________________
\subsection{Physical properties}\label{sec:pars}

Figure~\ref{fig:chaipars} shows a comparison between the physical parameters derived from the SED fitting for our new candidate members and the confirmed members of the Cha~I moving group. To achieve a homogeneous comparison, the SEDs of the confirmed members were constructed using the photometric data provided in Paper~I, and fitted to model photospheres as described in Sect.~\ref{sec:fits}. The best-fitting result was selected in each case, regardless of the $\dot{\chi}^2$ value, provided that the parameters were in good agreement with the known spectroscopic properties of the objects. In the final sample, the fits all had $\dot{\chi}^2<100$ and, in most cases, Bayesian probabilities larger than 50\% for the derived $T_{eff}$ and $A_V$ values. The  best-fitting results and physical parameters for the Cha~I kinematical members are summarized in Table~\ref{tab:mempars}. We remark that this table includes only those objects with enough datapoints to perform a fit. Hence, the sample of confirmed Cha~I members with derived physical parameters is far from complete: it includes only 39 stars. Furthermore, masses and ages were estimated for only 24 objects of this sample, as the rest of sources lie above the 1~Myr isochrone in the Hertzprung-Russell diagram. Similarly, the candidate sample includes only those objects classified as probable members in the histograms of the lower panels, while the possible member \object{ChaI-PM-4} is also considered in the histograms of the upper panels.

The effective temperature distributions for the new members and the confirmed members are shown in  the upper-left panel of Fig.~\ref{fig:chaipars}. Despite covering a broad range of effective temperature reaching the A-type domain ($2700\lesssim T_{eff}\lesssim10000$), M-type stars dominate the confirmed member sample, with only 25\% of K-type or earlier stars. The temperature range spanned by the candidate sample ($3400\lesssim T_{eff}\lesssim4200$) is much narrower than that covered by the confirmed member sample, and it seems to have a deficit of both early-type and M-type objects. Except for one source that is cooler (namely \object{ChaI-PM-48}), the candidates have effective temperatures between 3900 and 4200~K, corresponding to mid-K spectral types. In agreement with these temperature distributions, the objects in the candidate sample also span a much narrower luminosity range than the confirmed members, as seen in the upper-right panel of Fig.~\ref{fig:chaipars}. However, both luminosity distributions peak around $\log L/L_{\odot}\sim-0.5$. 

The tail of bright, hot stars seen in the confirmed member distributions is formed by relatively massive stars (compared with the candidate sample) lying above the youngest isochrone in the Hertzprung-Russell diagram, and hence very young and in a relatively early evolutionary stage according to their SED shapes. If these objects are not considered and we restrict our analysis to M- and K-type objects, the temperature and luminosity distributions of confirmed members and candidates are quite similar, although a trend towards slightly hotter effective temperatures and accordingly brighter luminosities is seen in the new candidates.

The corresponding mass and age distributions are shown in the lower panels of Fig.~\ref{fig:chaipars}. No masses and ages could be derived for the objects placed above the 1~Myr isochrone in the Hertzprung-Russell diagram, including the hottest and most luminous members discussed above. Therefore, the mass distribution of the Cha~I members does not match their effective temperature distribution, and is not expected to be representative of the whole cloud population. However, this is not a critical issue for our new members, because it affects only one object (\object{ChaI-PM-4}). The lower-left panel of Fig.~\ref{fig:chaipars} shows that the new probable members span a similar range in masses to the confirmed members for which this parameter could be estimated, but it tends to objects with higher masses, in agreement with the average higher effective temperatures. The lower-right panel of Fig.~\ref{fig:chaipars} shows that both samples have similar age distributions, as expected if they belong to the same population.

The physical parameters for the confirmed members of the Cha~II moving group were derived in the same way as explained above for the Cha~I members, and compared to the parameter values for the only new probable member, \object{ChaII-PM-11}. Of the 22 kinematic members identified in Paper~I, we were able to obtain physical parameters for 14 stars, which are summarized in Table~\ref{tab:mempars}. Masses were estimated for nine of these objects, while ages could be derived for eight sources. The rest of confirmed members lie above the 1~Myr isochrone in the Hertzprung-Russell diagram, and thus no masses or ages could be derived for them.

The physical parameters of the new probable member suggest that it is comparable to the lowest-mass (thus, coolest and lowest-luminosity) objects in the member sample. We note, however, that lower mass objects are known in Cha~II, but fall below the UCAC sensitivity and are thus excluded from our statistics.

%
%########################################################################
\section{Summary and conclusions}\label{sec:concl}

Based on proper motion and photometric information, we have identified 51 and 14 kinematic candidate members to the Cha~I and II moving groups, respectively, in an area of three degrees around each cloud. The SED analysis suggests that seventeen objects from the Cha~I sample and one from the Cha~II sample are likely young stars, and thus probable members of the moving groups. Another object from the Cha~I sample is located slightly above the 1~Myr isochrone in the Hertzprung-Russell diagram, and is classified as a possible member to the Cha~I moving group. The nature of the remaining candidates is more uncertain.

Our new probable members span a mass range $0.3\lesssim M/M_{\odot}\lesssim1.4$, and their physical properties are very similar to those of the confirmed members of both clouds included in UCAC. 
Thus, our results suggest that the stellar population associated to the current Cha~I cloud may be larger and more widespread than previously thought. In Cha~II, this is more uncertain, because only one object has been classified as a probable member. Similar off-cloud populations have been reported in other nearby low-mass star forming regions, like e.g. the Lupus clouds \citep{sparlup, speclup}. 

To find the answers to the questions raised by this work, spectroscopic confirmation of the nature of our new candidate members is needed, along with more accurate proper motions and more consistent age estimates. The {\em Gaia} mission of the European Space Agency, complemented with ongoing and future spectroscopic surveys, will provide this information with unprecedented precision for most of the stars in the sample studied here, allowing a much better understanding of the properties of the stellar population and the star formation history in the Chamaeleon clouds.

%########################################################################
 \begin{acknowledgements}
 
This work was funded by the Spanish MICINN through grant Consolider-CSD2006-00070. It also benefited from funding from the Spanish government through grants AYA2008-02156, AYA2010-21161-C02-02, and AYA2012-38897-C01-01, and from the Madrid regional government through grant PRICIT-S2009ESP-1496. A.~B. was co-funded under the Marie Curie Actions of the European Commission (FP7-COFUND). H.~B. is funded by the Ram\'on y Cajal Fellowship Program under grant number RYC-2009-04497. B.~L.~M. also thanks support from the Spanish government through project AYA2011-30147-C03-03.
 
This publication makes use of VOSA, developed under the Spanish Virtual Observatory project supported from the Spanish MEC through grant AYA2008-02156. It greatly benefited from the use of the SIMBAD database and VIZIER catalogue service, both operated at the CDS (Strasbourg, France). We used the VO-compliant tools Aladin, developed at the CDS, and TOPCAT, currently developed within the AstroGrid project. 

In addition, this publication makes use of data products from the Two Micron All Sky Survey (2MASS), which is a joint project of the University of Massachusetts and the Infrared Processing and Analysis Center/California Institute of Technology, funded by the US National Aeronautics and Space Administration and National Science Foundation; and from the Wide-field Infrared Survey Explorer (WISE), which is a joint project of the University of California, Los Angeles, and the Jet Propulsion Laboratory/California Institute of Technology, funded by the US National Aeronautics and Space Administration.

\end{acknowledgements}

%########################################################################
%########################################################################

\bibliographystyle{aa} % style aa.bst
\bibliography{references}

%########################################################################

%### TABLES

%_____________________________________________________________
%                              Table longer than a single page
%                                             and in landscape 
%  In the preamble, use:       \usepackage{longtable,lscape}
%-------------------------------------------------------------
%          All long tables have to be placed at the end, after
%                                        \end{thebibliography}
%
% In the text, at the place where the large table should appear
% add the command:                        
%\addtocounter{table}{1} 
% Tables counters will be well numbered.
%-------------------------------------------------------------

\longtabL{2}{
\begin{landscape}
\begin{longtable}{l c c c c c r r r r r c c c r}

% first head
\caption{Kinematical candidate members to the Cha~I and II moving groups}
\label{tab:cand}\\
    \hline \hline 
    \noalign{\smallskip}
  \multicolumn{1}{l}{Object} &
  \multicolumn{1}{c}{UCAC4} &
  \multicolumn{1}{c}{$\alpha$} &
  \multicolumn{1}{c}{$\delta$} &
  \multicolumn{1}{c}{$\mu_{\alpha}\cos\delta$} &
  \multicolumn{1}{c}{$\mu_{\delta}$} &
  \multicolumn{1}{c}{$B$} &
  \multicolumn{1}{c}{$V$} &
  \multicolumn{1}{c}{$g'$} &
  \multicolumn{1}{c}{$r'$} &
  \multicolumn{1}{c}{$i'$} &
  \multicolumn{1}{c}{$J$} &
  \multicolumn{1}{c}{$H$}  &
  \multicolumn{1}{c}{$Ks$} &
  \multicolumn{1}{c}{$A^{grid}_V$}  \\  
  \multicolumn{1}{l}{} &
  \multicolumn{1}{c}{} &
  \multicolumn{1}{c}{(hh mm ss)} &
  \multicolumn{1}{c}{(dd mm ss)} &
  \multicolumn{1}{c}{(mas/yr)} &
  \multicolumn{1}{c}{(mas/yr)} &
  \multicolumn{1}{c}{(mag)} &
  \multicolumn{1}{c}{(mag)} &
  \multicolumn{1}{c}{(mag)} &
  \multicolumn{1}{c}{(mag)} &
  \multicolumn{1}{c}{(mag)} &
  \multicolumn{1}{c}{(mag)} &
  \multicolumn{1}{c}{(mag)}  &
  \multicolumn{1}{c}{(mag)}  &
  \multicolumn{1}{c}{(mag)}  \\
      \noalign{\smallskip}\hline
      \noalign{\smallskip}
\endfirsthead

%second heads
\caption{(continued)}\\
   \hline\hline
   \noalign{\smallskip}
  \multicolumn{1}{l}{Object} &
  \multicolumn{1}{c}{UCAC4} &
  \multicolumn{1}{c}{$\alpha$} &
  \multicolumn{1}{c}{$\delta$} &
  \multicolumn{1}{c}{$\mu_{\alpha}\cos\delta$} &
  \multicolumn{1}{c}{$\mu_{\delta}$} &
  \multicolumn{1}{c}{$B$} &
  \multicolumn{1}{c}{$V$} &
  \multicolumn{1}{c}{$g'$} &
  \multicolumn{1}{c}{$r'$} &
  \multicolumn{1}{c}{$i'$} &
  \multicolumn{1}{c}{$J$} &
  \multicolumn{1}{c}{$H$}  &
  \multicolumn{1}{c}{$Ks$}  &
  \multicolumn{1}{c}{$A^{grid}_V$}  \\  
  \multicolumn{1}{l}{} &
  \multicolumn{1}{c}{} &
  \multicolumn{1}{c}{(hh mm ss)} &
  \multicolumn{1}{c}{(dd mm ss)} &
  \multicolumn{1}{c}{(mas/yr)} &
  \multicolumn{1}{c}{(mas/yr)} &
  \multicolumn{1}{c}{(mag)} &
  \multicolumn{1}{c}{(mag)} &
  \multicolumn{1}{c}{(mag)} &
  \multicolumn{1}{c}{(mag)} &
  \multicolumn{1}{c}{(mag)} &
  \multicolumn{1}{c}{(mag)} &
  \multicolumn{1}{c}{(mag)}  &
  \multicolumn{1}{c}{(mag)}  &
  \multicolumn{1}{c}{(mag)}  \\
      \noalign{\smallskip}\hline
      \noalign{\smallskip}
\endhead
\hline
\endfoot

% Start table
  \multicolumn{12}{c}{Cha~I moving group}\\
  \noalign{\smallskip}

  \object{ChaI-PM-1}  & 051-009778 & 11 24 19.03 & $-$79 51 48.7 & $-$23.9$\pm$1.2 &    1.4$\pm$1.1 & 13.09 & 11.67 & 12.37 & 11.15 & 10.56 &  8.97$\pm$0.02 &  8.22$\pm$0.05 &  8.05$\pm$0.03 & 0.24 \\
  \object{ChaI-PM-2}  & 052-010029 & 11 38 34.93 & $-$79 47 33.1 & $-$23.3$\pm$2.9 &    4.4$\pm$2.5 & 14.21 & 12.69 & 13.44 & 12.13 & 11.53 &  9.89$\pm$0.02 &  9.14$\pm$0.03 &  8.95$\pm$0.02 & 0.41 \\
  \object{ChaI-PM-3}  & 054-010728 & 11 36 28.16 & $-$79 22 20.2 & $-$16.0$\pm$2.9 &    0.7$\pm$1.4 & 13.21 & 11.88 & 12.51 & 11.41 & 10.95 &  9.49$\pm$0.02 &  8.82$\pm$0.05 &  8.65$\pm$0.02 & 0.00 \\
  \object{ChaI-PM-4}  & 054-010966 & 11 47 39.05 & $-$79 13 57.6 & $-$15.2$\pm$1.3 &    1.6$\pm$1.6 & 12.69 & 11.17 & 11.91 & 10.62 & 10.04 &  8.29$\pm$0.02 &  7.57$\pm$0.03 &  7.33$\pm$0.02 & 0.17 \\
  \object{ChaI-PM-5}  & 055-010397 & 11 19 35.35 & $-$79 11 25.4 & $-$17.0$\pm$0.8 &    0.7$\pm$1.0 & 12.23 & 10.59 & 11.38 & 10.11 &  9.20 &  7.43$\pm$0.03 &  6.59$\pm$0.03 &  6.35$\pm$0.02 & 0.71 \\
  \object{ChaI-PM-6}  & 055-010573 & 11 26 35.10 & $-$79 03 26.7 & $-$19.6$\pm$1.6 &    4.9$\pm$1.0 & 13.03 & 11.59 & 12.30 & 11.07 & 10.52 &  8.93$\pm$0.02 &  8.25$\pm$0.04 &  8.05$\pm$0.02 & 0.00 \\
  \object{ChaI-PM-7}  & 055-010743 & 11 33 56.53 & $-$79 10 08.1 & $-$15.2$\pm$1.4 & $-$0.5$\pm$1.4 & 14.15 & 12.76 & 13.43 & 12.30 & 11.76 & 10.15$\pm$0.02 &  9.52$\pm$0.02 &  9.33$\pm$0.02 & 0.00 \\
  \object{ChaI-PM-8}  & 056-010912 & 11 23 11.10 & $-$78 56 14.4 & $-$16.0$\pm$3.3 &    3.9$\pm$4.1 & 13.56 & 12.12 & 12.84 & 11.63 & 11.11 &  9.61$\pm$0.03 &  8.93$\pm$0.02 &  8.77$\pm$0.02 & 0.00 \\
  \object{ChaI-PM-9}  & 056-011976 & 11 55 08.09 & $-$78 50 28.1 & $-$18.1$\pm$3.6 &    2.4$\pm$1.7 & 13.72 & 12.20 & 12.95 & 11.68 & 11.27 &  9.70$\pm$0.02 &  9.07$\pm$0.02 &  8.86$\pm$0.02 & 0.00 \\
%  \object{ChaI-PM-10} & 057-011234 & 11 45 13.83 & $-$78 36 58.6 & $-$19.8$\pm$1.1 &    2.1$\pm$1.0 &  9.95 &  9.20 & 10.55 &  9.29 &  8.24 &  6.62$\pm$0.03 &  6.11$\pm$0.05 &  5.95$\pm$0.03 & 0.11 \\%
  \object{ChaI-PM-11} & 058-009871 & 10 31 21.57 & $-$78 27 48.1 & $-$21.3$\pm$1.6 &    0.0$\pm$1.4 & 13.17 & 11.90 & 12.52 & 11.44 & 10.91 &  9.39$\pm$0.02 &  8.78$\pm$0.07 &  8.55$\pm$0.02 & 0.35 \\
  \object{ChaI-PM-12} & 059-010277 & 10 57 49.82 & $-$78 20 39.3 & $-$17.6$\pm$1.7 &    4.0$\pm$1.6 & 15.33 & 13.80 & 14.54 & 13.22 & 12.63 & 10.81$\pm$0.02 & 10.15$\pm$0.02 &  9.92$\pm$0.02 & 0.00 \\
  \object{ChaI-PM-13} & 059-010308 & 10 59 38.22 & $-$78 22 42.1 & $-$21.6$\pm$1.1 &    5.6$\pm$1.4 & 12.78 & 11.07 & 11.92 & 10.40 &  9.67 &  7.75$\pm$0.02 &  6.95$\pm$0.05 &  6.69$\pm$0.02 & 1.08 \\
  \object{ChaI-PM-14} & 059-010316 & 11 00 15.59 & $-$78 14 10.2 & $-$15.2$\pm$1.5 & $-$2.2$\pm$1.5 & 15.46 & 13.71 & 14.59 & 12.97 & 12.13 &  9.96$\pm$0.03 &  9.10$\pm$0.02 &  8.83$\pm$0.02 & 1.73 \\
%  \object{ChaI-PM-15} & 060-012395 & 11 59 48.08 & $-$78 11 45.0 & $-$23.0$\pm$2.4 & $-$0.7$\pm$8.0 & 10.39 &  9.27 & 10.53 &  9.04 &  8.12 &  6.59$\pm$0.03 &  5.98$\pm$0.04 &  5.85$\pm$0.02 & 0.59 \\%
  \object{ChaI-PM-16} & 062-010209 & 10 48 26.64 & $-$77 39 19.2 & $-$19.5$\pm$1.0 &    4.7$\pm$1.2 & 13.04 & 11.45 & 12.25 & 10.84 & 10.14 &  8.26$\pm$0.03 &  7.54$\pm$0.05 &  7.30$\pm$0.02 & 0.00 \\
  \object{ChaI-PM-17} & 062-010325 & 10 56 18.54 & $-$77 44 53.0 & $-$18.0$\pm$1.5 & $-$1.1$\pm$1.2 & 14.56 & 12.41 & 13.43 & 11.50 &  9.76 &  6.92$\pm$0.03 &  5.84$\pm$0.04 &  5.45$\pm$0.02 & 0.00 \\
  \object{ChaI-PM-18} & 063-011258 & 11 17 57.98 & $-$77 28 39.2 & $-$17.3$\pm$5.3 &    2.7$\pm$4.2 & 17.82 & 16.42 & 16.99 & 15.6  & 13.87 & 11.53$\pm$0.03 & 10.88$\pm$0.02 & 10.64$\pm$0.03 & 0.00 \\
  \object{ChaI-PM-19} & 063-012349 & 11 46 09.78 & $-$77 27 54.3 & $-$16.9$\pm$3.8 &    3.7$\pm$1.2 & 13.58 & 12.04 & 12.80 & 11.51 & 11.03 &  9.55$\pm$0.02 &  8.91$\pm$0.03 &  8.71$\pm$0.02 & 0.00 \\
  \object{ChaI-PM-20} & 064-010772 & 11 12 05.15 & $-$77 12 40.7 & $-$15.1$\pm$1.5 & $-$2.2$\pm$1.7 & 16.51 & 14.27 & 15.42 & 13.21 & 12.13 &  9.36$\pm$0.02 &  8.21$\pm$0.04 &  7.85$\pm$0.03 & 0.00 \\
  \object{ChaI-PM-21} & 064-011840 & 11 43 48.52 & $-$77 16 39.8 & $-$20.9$\pm$1.2 &    5.6$\pm$1.2 & 14.01 & 12.65 & 13.31 & 12.17 & 11.67 & 10.19$\pm$0.02 &  9.58$\pm$0.02 &  9.41$\pm$0.02 & 0.00 \\
  \object{ChaI-PM-22} & 065-010266 & 10 18 44.30 & $-$77 01 05.3 & $-$16.0$\pm$1.2 &    2.0$\pm$2.1 & 13.42 & 12.06 & 12.71 & 11.58 & 11.00 &  9.38$\pm$0.03 &  8.78$\pm$0.07 &  8.60$\pm$0.02 & 0.00 \\
  \object{ChaI-PM-23} & 067-010946 & 10 31 57.71 & $-$76 44 25.5 & $-$19.4$\pm$1.3 &    4.6$\pm$1.2 & 13.73 & 12.15 & 12.92 & 11.58 & 11.01 &  9.35$\pm$0.02 &  8.68$\pm$0.05 &  8.46$\pm$0.02 & 0.16 \\
  \object{ChaI-PM-24} & 067-011535 & 10 50 50.59 & $-$76 44 48.9 & $-$14.9$\pm$1.2 &    1.8$\pm$1.5 & 13.28 & 11.97 & 12.58 & 11.48 & 11.01 &  9.43$\pm$0.02 &  8.85$\pm$0.02 &  8.62$\pm$0.02 & 0.00 \\
  \object{ChaI-PM-25} & 067-012242 & 11 22 26.40 & $-$76 46 08.7 & $-$19.1$\pm$1.3 &    3.6$\pm$1.4 & 13.92 & 12.39 & 13.15 & 11.81 & 11.28 &  9.64$\pm$0.02 &  8.93$\pm$0.04 &  8.73$\pm$0.02 & 0.00 \\
  \object{ChaI-PM-26} & 067-013090 & 11 46 11.39 & $-$76 37 26.0 & $-$19.1$\pm$3.0 &    3.3$\pm$1.1 & 13.09 & 11.64 & 12.36 & 11.14 & 10.69 &  9.20$\pm$0.03 &  8.59$\pm$0.05 &  8.39$\pm$0.03 & 0.00 \\
  \object{ChaI-PM-27} & 068-011668 & 10 39 31.02 & $-$76 33 41.0 & $-$21.0$\pm$1.5 &    4.3$\pm$1.8 & 12.94 & 11.51 & 12.22 & 10.99 & 10.47 &  8.90$\pm$0.03 &  8.2 $\pm$0.04 &  7.99$\pm$0.04 & 0.00 \\
  \object{ChaI-PM-28} & 068-012962 & 11 21 30.81 & $-$76 33 35.2 & $-$15.4$\pm$2.1 & $-$1.0$\pm$2.2 & 16.80 & 15.07 & 15.92 & 14.32 & 13.12 & 10.90$\pm$0.02 & 10.15$\pm$0.02 &  9.88$\pm$0.02 & 0.00 \\
  \object{ChaI-PM-29} & 069-012749 & 11 12 00.38 & $-$76 14 10.5 & $-$15.0$\pm$1.6 &    2.0$\pm$1.7 & 14.78 & 13.10 & 13.94 & 12.43 & 11.70 &  9.52$\pm$0.03 &  8.65$\pm$0.04 &  8.36$\pm$0.02 & 0.00 \\
  \object{ChaI-PM-30} & 069-013658 & 11 35 52.56 & $-$76 19 01.0 & $-$20.0$\pm$1.6 &    1.2$\pm$1.6 & 14.50 & 13.12 & 13.79 & 12.62 & 12.10 & 10.50$\pm$0.03 &  9.84$\pm$0.02 &  9.67$\pm$0.02 & 0.00 \\
  \object{ChaI-PM-31} & 071-013120 & 10 52 51.59 & $-$75 54 37.3 & $-$15.5$\pm$1.4 &    4.2$\pm$1.4 & 13.59 & 11.88 & 12.71 & 11.22 & 10.51 &  8.65$\pm$0.02 &  7.84$\pm$0.05 &  7.61$\pm$0.03 & 0.00 \\
  \object{ChaI-PM-32} & 071-015025 & 11 37 25.57 & $-$75 52 05.1 & $-$21.2$\pm$2.1 &    5.6$\pm$2.1 & 15.61 & 14.13 & 14.83 & 13.52 & 12.79 & 10.87$\pm$0.02 & 10.32$\pm$0.02 & 10.06$\pm$0.02 & 0.00 \\
  \object{ChaI-PM-33} & 071-015061 & 11 39 12.53 & $-$75 50 46.7 & $-$15.3$\pm$2.2 & $-$1.1$\pm$2.1 & 15.76 & 14.07 & 14.88 & 13.46 & 12.78 & 10.88$\pm$0.02 & 10.13$\pm$0.02 &  9.91$\pm$0.02 & 0.00 \\
  \object{ChaI-PM-34} & 071-015244 & 11 45 05.21 & $-$75 48 48.9 & $-$20.8$\pm$1.1 &    5.4$\pm$1.1 & 12.94 & 11.39 & 12.12 & 10.83 & 10.31 &  8.63$\pm$0.03 &  7.96$\pm$0.02 &  7.78$\pm$0.03 & 0.00 \\
  \object{ChaI-PM-35} & 071-015400 & 11 48 49.56 & $-$75 57 55.0 & $-$16.5$\pm$1.0 &    3.5$\pm$1.1 & 12.15 & 10.53 & 11.41 &  9.98 &  9.31 &  7.54$\pm$0.03 &  6.77$\pm$0.04 &  6.58$\pm$0.03 & 0.00 \\
%  \object{ChaI-PM-36} & 072-014984 & 11 29 12.63 & $-$75 46 26.4 & $-$22.6$\pm$1.5 & $-$1.9$\pm$1.5 & 14.35 & 12.92 & 13.64 & 12.32 & 11.65 &  9.82$\pm$0.03 &  9.12$\pm$0.02 &  8.88$\pm$0.02 & 0.00 \\
  \object{ChaI-PM-37} & 072-015250 & 11 36 33.52 & $-$75 47 14.3 & $-$15.7$\pm$2.3 &    0.0$\pm$2.1 & 16.75 & 14.86 & 15.81 & 14.02 & 13.04 & 10.65$\pm$0.02 &  9.73$\pm$0.02 &  9.45$\pm$0.02 & 0.00 \\
  \object{ChaI-PM-38} & 072-015687 & 11 47 56.70 & $-$75 36 31.7 & $-$17.5$\pm$1.1 &    1.2$\pm$1.1 & 12.55 & 10.63 & 11.56 &  9.97 &  9.02 &  6.97$\pm$0.02 &  6.08$\pm$0.03 &  5.80$\pm$0.02 & 0.00 \\
  \object{ChaI-PM-39} & 073-015115 & 11 14 26.21 & $-$75 25 10.3 & $-$23.0$\pm$1.1 &    5.6$\pm$1.3 & 12.22 & 10.84 & 11.53 & 10.44 &  9.96 &  8.38$\pm$0.03 &  7.75$\pm$0.04 &  7.56$\pm$0.02 & 0.00 \\
  \object{ChaI-PM-40} & 073-015796 & 11 30 01.54 & $-$75 29 14.6 & $-$15.8$\pm$1.5 &    2.9$\pm$1.5 & 13.93 & 12.35 & 13.13 & 11.75 & 11.09 &  9.29$\pm$0.02 &  8.60$\pm$0.04 &  8.32$\pm$0.04 & 0.00 \\
  \object{ChaI-PM-41} & 073-015916 & 11 32 33.52 & $-$75 35 01.1 & $-$18.8$\pm$1.0 &    2.4$\pm$1.0 & 12.15 & 10.58 & 11.32 &  9.98 &  9.33 &  7.59$\pm$0.02 &  6.83$\pm$0.02 &  6.59$\pm$0.02 & 0.00 \\
  \object{ChaI-PM-42} & 073-015951 & 11 33 24.12 & $-$75 32 02.2 & $-$16.2$\pm$1.9 & $-$0.7$\pm$2.0 & 15.33 & 13.73 & 14.50 & 13.07 & 12.44 & 10.67$\pm$0.02 &  9.94$\pm$0.03 &  9.73$\pm$0.02 & 0.00 \\
  \object{ChaI-PM-43} & 073-016070 & 11 36 26.68 & $-$75 31 42.5 & $-$17.3$\pm$1.7 &    5.4$\pm$1.7 & 15.09 & 13.56 & 14.30 & 12.98 & 12.30 & 10.47$\pm$0.02 &  9.69$\pm$0.02 &  9.50$\pm$0.02 & 0.00 \\
%  \object{ChaI-PM-44} & 074-014708 & 11 08 53.26 & $-$75 19 37.4 & $-$22.3$\pm$1.7 &    1.7$\pm$1.7 & 14.61 & 13.13 & 13.87 & 12.50 & 11.70 &  9.74$\pm$0.03 &  9.03$\pm$0.05 &  8.80$\pm$0.02 & 0.00 \\
  \object{ChaI-PM-45} & 074-015749 & 11 30 52.27 & $-$75 22 44.6 & $-$17.9$\pm$1.3 &    1.1$\pm$1.3 & 13.61 & 12.37 & 12.96 & 11.90 & 11.45 &  9.97$\pm$0.02 &  9.30$\pm$0.02 &  9.18$\pm$0.02 & 0.00 \\
  \object{ChaI-PM-46} & 076-014623 & 10 47 36.41 & $-$74 54 33.3 & $-$14.8$\pm$1.6 & $-$0.5$\pm$1.6 & 14.77 & 13.25 & 13.97 & 12.70 & 12.18 & 10.53$\pm$0.02 &  9.92$\pm$0.02 &  9.72$\pm$0.02 & 0.00 \\
  \object{ChaI-PM-47} & 076-015051 & 10 56 14.89 & $-$74 50 45.5 & $-$14.8$\pm$1.4 & $-$0.8$\pm$1.4 & 13.37 & 11.64 & 12.49 & 11.01 & 10.19 &  8.32$\pm$0.02 &  7.52$\pm$0.04 &  7.22$\pm$0.02 & 0.00 \\
  \object{ChaI-PM-48} & 076-015563 & 11 05 51.66 & $-$74 58 39.6 & $-$20.9$\pm$2.6 &    1.9$\pm$2.6 & 16.52 & 15.03 & 15.88 & 14.27 & 13.28 & 11.10$\pm$0.02 & 10.37$\pm$0.02 & 10.14$\pm$0.02 & 0.00 \\
%  \object{ChaI-PM-49} & 076-016832 & 11 27 57.80 & $-$74 50 43.5 & $-$16.3$\pm$1.1 &    3.0$\pm$1.1 & 11.89 & 10.17 & 11.02 &  9.61 &  8.75 &  6.96$\pm$0.02 &  6.07$\pm$0.02 &  5.86$\pm$0.02 & 0.00 \\%
  \object{ChaI-PM-50} & 076-016843 & 11 28 14.58 & $-$74 53 08.5 & $-$21.7$\pm$1.6 & $-$0.4$\pm$1.1 & 12.27 & 10.60 & 11.43 & 10.01 &  9.36 &  7.68$\pm$0.04 &  6.85$\pm$0.04 &  6.58$\pm$0.02 & 0.00 \\
  \object{ChaI-PM-51} & 077-015492 & 11 03 49.69 & $-$74 39 48.7 & $-$18.3$\pm$1.8 &    4.9$\pm$1.4 & 12.66 & 11.17 & 11.90 & 10.65 & 10.12 &  8.46$\pm$0.02 &  7.80$\pm$0.05 &  7.57$\pm$0.03 & 0.00 \\
  \object{ChaI-PM-52} & 077-016041 & 11 12 31.20 & $-$74 45 26.7 & $-$17.8$\pm$2.9 &    0.7$\pm$2.9 & 16.78 & 15.04 & 15.86 & 14.35 & 13.07 & 10.85$\pm$0.03 & 10.05$\pm$0.03 &  9.84$\pm$0.02 & 0.00 \\
  \object{ChaI-PM-53} & 078-018302 & 11 21 10.28 & $-$74 28 56.6 & $-$15.2$\pm$1.4 & $-$0.4$\pm$2.6 & 13.95 & 12.56 & 13.23 & 12.08 & 11.60 & 10.11$\pm$0.03 &  9.44$\pm$0.02 &  9.29$\pm$0.02 & 0.00 \\
%  \object{ChaI-PM-54} & 078-018348 & 11 22 01.47 & $-$74 27 27.3 & $-$17.4$\pm$1.1 &    0.3$\pm$1.2 & 12.21 & 10.37 & 11.34 &  9.81 &  8.76 &  6.87$\pm$0.02 &  6.01$\pm$0.05 &  5.70$\pm$0.02 & 0.00 \\%
  \object{ChaI-PM-55} & 078-018479 & 11 24 47.80 & $-$74 35 00.8 & $-$15.2$\pm$1.2 &    3.1$\pm$1.2 & 13.19 & 11.65 & 12.40 & 11.12 & 10.63 &  9.03$\pm$0.02 &  8.41$\pm$0.05 &  8.21$\pm$0.03 & 0.00 \\
  \object{ChaI-PM-56} & 079-016934 & 10 57 31.50 & $-$74 23 06.8 & $-$16.1$\pm$1.5 & $-$1.2$\pm$1.5 & 13.57 & 12.16 & 12.84 & 11.69 & 11.20 &  9.64$\pm$0.03 &  8.99$\pm$0.02 &  8.78$\pm$0.02 & 0.00 \\
  \object{ChaI-PM-57} & 079-017285 & 11 03 53.92 & $-$74 19 25.4 & $-$19.7$\pm$1.5 &    1.7$\pm$1.5 & 13.87 & 12.59 & 13.19 & 12.19 & 11.76 & 10.23$\pm$0.03 &  9.56$\pm$0.02 &  9.41$\pm$0.02 & 0.00 \\

  \noalign{\smallskip}\hline\noalign{\smallskip}
  \multicolumn{12}{c}{Cha~II moving group}\\
  \noalign{\smallskip}

  \object{ChaII-PM-1}  & 050-011288 & 12 47 25.39 & $-$80 08 01.5 & $-$19.0$\pm$2.1 & $-$10.7$\pm$4.6 & 17.67 & 15.80 & 16.74 & 14.89 & 13.75 & 11.21$\pm$0.02 & 10.47$\pm$0.02 & 10.14$\pm$0.02 & 2.81 \\
  \object{ChaII-PM-2}  & 054-012268 & 12 54 34.93 & $-$79 23 11.0 & $-$17.6$\pm$1.1 &  $-$2.2$\pm$1.1 & 13.83 & 12.04 & 12.95 & 11.31 & 10.40 &  8.30$\pm$0.03 &  7.50$\pm$0.03 &  7.22$\pm$0.02 & 0.92 \\
  \object{ChaII-PM-3}  & 055-012803 & 12 54 27.89 & $-$79 07 12.6 & $-$16.5$\pm$1.5 &  $-$3.0$\pm$1.5 & 16.11 & 13.66 & 14.89 & 12.62 & 11.13 &  8.27$\pm$0.03 &  7.11$\pm$0.04 &  6.65$\pm$0.02 & 2.80 \\
%  \object{ChaII-PM-4}  & 055-013876 & 13 29 27.82 & $-$79 07 03.7 & $-$23.7$\pm$1.2 &  $-$4.7$\pm$1.1 & 11.10 &  9.46 & 10.25 &  8.76 &  7.03 &  2.36$\pm$0.27 &  1.46$\pm$0.29 &  1.04$\pm$0.31 & 3.32 \\
%  \object{ChaII-PM-5}  & 059-014997 & 13 14 02.48 & $-$78 14 42.8 & $-$29.7$\pm$1.2 &  $-$6.8$\pm$1.8 & 11.00 &  9.58 & 10.18 &  9.01 &  8.08 &  6.31$\pm$0.02 &  5.56$\pm$0.02 &  5.26$\pm$0.02 & 0.00 \\
  \object{ChaII-PM-6}  & 060-014727 & 13 04 33.12 & $-$78 06 07.9 & $-$16.0$\pm$2.2 &  $-$2.8$\pm$2.0 & 15.56 & 13.74 & 14.68 & 12.97 & 12.08 &  9.86$\pm$0.02 &  8.98$\pm$0.03 &  8.65$\pm$0.02 & 1.18 \\
  \object{ChaII-PM-7}  & 061-014507 & 13 09 04.31 & $-$77 54 39.0 & $-$17.2$\pm$1.5 &  $-$5.7$\pm$1.5 & 15.02 & 13.37 & 14.22 & 12.66 & 11.87 &  9.87$\pm$0.03 &  9.06$\pm$0.03 &  8.82$\pm$0.02 & 0.25 \\
  \object{ChaII-PM-8}  & 063-014624 & 13 07 19.79 & $-$77 25 41.1 & $-$19.7$\pm$1.8 &  $-$2.3$\pm$1.8 & 15.77 & 13.83 & 14.81 & 12.98 & 12.07 &  9.75$\pm$0.02 &  8.86$\pm$0.03 &  8.51$\pm$0.02 & 0.00 \\
  \object{ChaII-PM-9}  & 066-014936 & 12 38 09.43 & $-$76 51 13.7 & $-$24.9$\pm$2.3 &  $-$5.8$\pm$2.3 & 15.42 & 13.56 & 14.51 & 12.48 &  9.82 &  6.13$\pm$0.02 &  5.12$\pm$0.06 &  4.67$\pm$0.04 & 0.00 \\
  \object{ChaII-PM-10} & 067-015388 & 12 52 51.72 & $-$76 45 34.4 & $-$22.0$\pm$1.9 &  $-$5.9$\pm$1.9 & 14.50 & 12.76 & 13.59 & 12.04 & 11.27 &  9.24$\pm$0.02 &  8.46$\pm$0.04 &  8.18$\pm$0.03 & 0.00 \\
  \object{ChaII-PM-11} & 067-015927 & 13 09 26.15 & $-$76 40 13.8 & $-$23.0$\pm$4.6 &  $-$8.2$\pm$4.5 & 16.75 & 15.10 & 15.87 & 14.45 & 13.16 & 11.03$\pm$0.02 & 10.26$\pm$0.02 & 10.05$\pm$0.02 & 0.00 \\
  \object{ChaII-PM-12} & 067-016250 & 13 15 34.41 & $-$76 47 13.5 & $-$20.2$\pm$2.0 &  $-$4.9$\pm$1.3 & 12.85 & 11.12 & 11.99 & 10.48 &  9.76 &  8.01$\pm$0.02 &  7.15$\pm$0.03 &  6.94$\pm$0.02 & 0.00 \\
  \object{ChaII-PM-13} & 072-019687 & 13 10 16.01 & $-$75 42 18.7 & $-$25.5$\pm$3.3 &  $-$2.1$\pm$3.3 & 15.55 & 13.81 & 14.63 & 13.01 & 10.76 &  8.09$\pm$0.03 &  7.12$\pm$0.03 &  6.75$\pm$0.02 & 0.00 \\
  \object{ChaII-PM-14} & 073-021942 & 13 23 10.45 & $-$75 30 40.1 & $-$18.0$\pm$1.9 &  $-$9.0$\pm$1.0 & 12.57 & 10.91 & 11.72 & 10.16 &  8.04 &  5.06$\pm$0.03 &  4.04$\pm$0.23 &  3.73$\pm$0.28 & 0.00 \\
%  \object{ChaII-PM-15} & 076-021082 & 12 42 33.25 & $-$74 58 44.5 & $-$28.5$\pm$1.6 &  $-$2.4$\pm$1.1 & 11.90 & 10.22 & 11.00 &  9.63 &  8.81 &  6.95$\pm$0.02 &  6.17$\pm$0.03 &  5.93$\pm$0.02 & 0.00 \\
%  \object{ChaII-PM-16} & 077-023922 & 13 18 31.54 & $-$74 45 50.7 & $-$19.7$\pm$1.4 &  $-$4.7$\pm$1.2 & 12.13 & 10.45 & 11.27 & 10.01 &  9.06 &  7.42$\pm$0.02 &  6.61$\pm$0.02 &  6.35$\pm$0.02 & 0.00 \\
  \object{ChaII-PM-17} & 078-025497 & 13 07 45.15 & $-$74 27 25.9 & $-$20.0$\pm$2.4 &  $-$4.8$\pm$2.4 & 13.66 & 11.96 & 12.78 & 11.33 & 10.18 &  8.32$\pm$0.02 &  7.45$\pm$0.04 &  7.15$\pm$0.02 & 0.00 \\
  \object{ChaII-PM-18} & 079-024691 & 12 58 38.11 & $-$74 17 11.2 & $-$25.6$\pm$8.3 &  $-$8.0$\pm$8.5 & 12.46 & 10.76 & 11.81 & 10.18 &  8.07 &  5.76$\pm$0.03 &  4.91$\pm$0.02 &  4.48$\pm$0.02 & 0.00 \\

  \noalign{\smallskip}
  \hline
\end{longtable}

% \begin{flushleft}
% {\bf Notes.} \\
%      (1) 
% \end{flushleft}

\end{landscape}
}
  % Cham I & II 

%_____________________________________________________________
%                                 A rotated Table in landscape  
%  In the preamble, use:   \usepackage{lscape}
%-------------------------------------------------------------
\begin{landscape}
\begin{table}
\label{tab:simbad}
\caption{Information available in SIMBAD for Chamaeleon~I and II new kinematical candidate members}
  \begin{center}
\begin{tabular}{ l l l c c c c c l l c}
\hline
\hline\noalign{\smallskip}
  \multicolumn{1}{l}{} &
  \multicolumn{1}{l}{} &
  \multicolumn{1}{l}{} &
  \multicolumn{2}{c}{UCAC4} &
  \multicolumn{2}{c}{Tycho-2} &
  \multicolumn{1}{c}{} &
  \multicolumn{1}{c}{} &  
  \multicolumn{1}{c}{} &
  \multicolumn{1}{c}{} \\
  \multicolumn{1}{l}{Object} &
  \multicolumn{1}{l}{UCAC4} &
  \multicolumn{1}{l}{SIMBAD Name} &
  \multicolumn{1}{c}{$\mu_{\alpha}\cos\delta$} &
  \multicolumn{1}{c}{$\mu_{\delta}$} &
  \multicolumn{1}{c}{$\mu_{\alpha}\cos\delta$} &
  \multicolumn{1}{c}{$\mu_{\delta}$} &
  \multicolumn{1}{c}{$V_r$} &
  \multicolumn{1}{c}{SpT} &  
  \multicolumn{1}{c}{References$^{\mathrm{a}}$} &
  \multicolumn{1}{c}{Remarks$^{\mathrm{b}}$} \\
  \multicolumn{1}{l}{} &
  \multicolumn{1}{l}{} &
  \multicolumn{1}{l}{} &
  \multicolumn{1}{c}{(mas/yr)} &
  \multicolumn{1}{c}{(mas/yr)} &
  \multicolumn{1}{c}{(mas/yr)} &
  \multicolumn{1}{c}{(mas/yr)} &
  \multicolumn{1}{c}{(km/s)} &
  \multicolumn{1}{c}{} &
  \multicolumn{1}{c}{} &
  \multicolumn{1}{c}{} \\
\noalign{\smallskip}\hline\noalign{\smallskip}
  \multicolumn{11}{c}{Chamaeleon~I candidates}\\
  \noalign{\smallskip}
\object{ChaI-PM-1}   & \object{051-009778} & \object{TYC~9419-417-1}	      & $-$23.9$\pm$1.2 &    1.4$\pm$1.1 &  $-$18.0$\pm$3.8  &    3.7$\pm$3.6  &	      &       & T    & \\
\object{ChaI-PM-3}   & \object{054-010728} & \object{TYC~9419-427-1}	      & $-$16.0$\pm$2.9 &    0.7$\pm$1.4 &      0.0$\pm$3.8  & $-$4.2$\pm$3.7  &	      &       & T    & \\
\object{ChaI-PM-4}   & \object{054-010966} & \object{TYC~9419-338-1}	      & $-$15.2$\pm$1.3 &    1.6$\pm$1.6 &   $-$8.4$\pm$2.7  &    4.4$\pm$2.6  &	      &       & T    & \\
\object{ChaI-PM-5}   & \object{055-010397} & \object{TYC~9418-388-1}	      & $-$17.0$\pm$0.8 &    0.7$\pm$1.0 &  $-$16.7$\pm$2.6  &    3.7$\pm$2.4  &	      &       & T    & \\
\object{ChaI-PM-6}   & \object{055-010573} & \object{TYC~9419-571-1}          & $-$19.6$\pm$1.6 &    4.9$\pm$1.0 &  $-$16.9$\pm$3.6  &    5.3$\pm$3.5  &              &       & T    & \\
\object{ChaI-PM-13}  & \object{059-010308} & \object{2MASS~J10593816-7822421} & $-$21.6$\pm$1.1 &    5.6$\pm$1.4 &  $-$26.7$\pm$5.4  &   15.5$\pm$5.2  & 11.0$\pm$0.7 &       & T, R & \\
\object{ChaI-PM-20}  & \object{064-010772} & \object{2MASS~J11120518-7712408} & $-$15.1$\pm$1.5 & $-$2.2$\pm$1.7 &	             &	               &              &       &      & \\
\object{ChaI-PM-24}  & \object{067-011535} & \object{TYC~9410-196-1}	      & $-$14.9$\pm$1.2 &    1.8$\pm$1.5 &  $-$13.8$\pm$5.5  &   10.3$\pm$5.3  &	      &       & T    & \\
\object{ChaI-PM-26}  & \object{067-013090} & \object{TYC~9411-1199-1}	      & $-$19.1$\pm$3.0 &    3.3$\pm$1.1 &  $-$30.8$\pm$3.7  &    3.8$\pm$3.6  &	      &       & T    & \\
\object{ChaI-PM-27}  & \object{068-011668} & \object{TYC~9397-1013-1}	      & $-$21.0$\pm$1.5 &    4.3$\pm$1.8 &  $-$14.7$\pm$5.5  &   15.6$\pm$5.3  &	      &       & T    & \\
\object{ChaI-PM-35}  & \object{071-015400} & \object{2MASS~J11484956-7557548} & $-$16.5$\pm$1.0 &    3.5$\pm$1.1 &  $-$18.3$\pm$2.1  &    3.8$\pm$2.0  &	      &       & T    & \\
\object{ChaI-PM-38}  & \object{072-015687} & \object{2MASS~J11475667-7536316} & $-$17.5$\pm$1.1 &    1.2$\pm$1.1 &  $-$21.4$\pm$2.9  &    4.1$\pm$2.4  &	      &       & T    & \\
\object{ChaI-PM-39}  & \object{073-015115} & \object{TYC 9410-603-1}	      & $-$23.0$\pm$1.1 &    5.6$\pm$1.3 &  $-$24.8$\pm$3.1  &   12.5$\pm$3.0  &	      &       & T    & \\
\object{ChaI-PM-41}  & \object{073-015916} & \object{2MASS~J11323348-7535011} & $-$18.8$\pm$1.0 &    2.4$\pm$1.0 &  $-$17.6$\pm$1.7  &    6.0$\pm$1.7  &	      &       & T    & \\
\object{ChaI-PM-50}  & \object{076-016843} & \object{2MASS~J11281458-7453084} & $-$21.7$\pm$1.6 & $-$0.4$\pm$1.1 &  $-$29.2$\pm$3.1  & $-$0.8$\pm$3.0  &	      &       & T    & \\
\object{ChaI-PM-51}  & \object{077-015492} & \object{TYC~9224-2496-1}	      & $-$18.3$\pm$1.8 &    4.9$\pm$1.4 &  $-$17.1$\pm$3.5  &    8.2$\pm$3.3  &	      &       & T    & \\
\object{ChaI-PM-55}  & \object{078-018479} & \object{TYC~9237-1121-1}	      & $-$15.2$\pm$1.2 &    3.1$\pm$1.2 &  $-$10.3$\pm$5.7  &    7.0$\pm$5.5  &	      &       & T    & \\

\noalign{\smallskip}\hline\noalign{\smallskip}
  \multicolumn{11}{c}{Chamaeleon~II candidates}\\
  \noalign{\smallskip}
\object{ChaII-PM-9}  & \object{066-014936} & \object{IRAS~12348-7634}	      & $-$24.9$\pm$2.3 & $-$5.8$\pm$2.3 & 	 	     &  	       &  	      &       &      & \\
\object{ChaII-PM-12} & \object{067-016250} & \object{TYC~9413-205-1}	      & $-$20.2$\pm$2.0 & $-$4.9$\pm$1.3 &  $-$18.0$\pm$2.6  &  $-$8.7$\pm$2.5 &  	      &       & T    & \\
\object{ChaII-PM-14} & \object{073-021942} & \object{IRAS~13191-7515}	      & $-$18.0$\pm$1.9 & $-$9.0$\pm$1.0 &  $-$19.7$\pm$2.1  & $-$10.8$\pm$2.0 &  	      &       & T    & \\
\object{ChaII-PM-18} & \object{079-024691} & \object{CR~Mus}		      & $-$25.6$\pm$8.3 & $-$8.0$\pm$8.5 & 	 	     &  	       &  	      & M4e   &      & bkg\\

  \noalign{\smallskip}\hline\noalign{\smallskip}
\end{tabular}
  \end{center}

\begin{flushleft}
{\bf Notes.} \\
     $^{\mathrm{a}}$ T$=$Tycho-2 \citep{hog2000}; R$=$RAVE \citep{siebert2011} \\  
     $^{\mathrm{b}}$ bkg$=$reported background object \citep{feigelson2003,luhman2004}%;  nm$=$non-member from this work 
      \\
     %$^{\mathrm{c}}$ The Tycho-2 proper motions are in better agreement with the median of the Cha~I moving group.\\
\end{flushleft}

\end{table}
\end{landscape}
%----------------------------------------------------------
  % Cham I & II 

%_____________________________________________________________
%        Some tables or figures are in the printed version and
%                      some are only in the electronic version
%-------------------------------------------------------------
%
% Leave all the tables or figures in the text, at their right place 
% and use the commands \onlfig{}{} and \onltab{}{}. These elements
% will be automatically placed at the end, in the section
% Online material.

%----------------------------------------------------------
% Table 4 available electronically only
\onltab{4}{
\longtabL{4}{
\begin{longtable}{l l r r r}

% first head
\caption{Compiled photometry for Cha~I and II candidate members.} \label{tab:phot} \\
\hline 
\hline  
\noalign{\smallskip}
  \multicolumn{1}{l}{Name} &
  \multicolumn{1}{l}{Filter$^{\mathrm{a}}$} &
  \multicolumn{1}{c}{$\lambda_{eff}$} &
  \multicolumn{1}{c}{Flux} &
  \multicolumn{1}{c}{Flux error$^{\mathrm{b}}$}  \\
  \multicolumn{1}{l}{} &
  \multicolumn{1}{c}{} &
  \multicolumn{1}{c}{(\AA)} &
  \multicolumn{1}{c}{(erg/cm$^2$/s/\AA)} &
  \multicolumn{1}{c}{(erg/cm$^2$/s/\AA)}  \\
\noalign{\smallskip}\hline
\noalign{\smallskip}
\endfirsthead

%second heads
\caption{(continued)}\\
\hline 
\hline  
\noalign{\smallskip}
  \multicolumn{1}{l}{Name} &
  \multicolumn{1}{l}{Filter$^{\mathrm{a}}$} &
  \multicolumn{1}{c}{$\lambda_{eff}$} &
  \multicolumn{1}{c}{Flux} &
  \multicolumn{1}{c}{Flux error$^{\mathrm{b}}$}  \\
  \multicolumn{1}{l}{} &
  \multicolumn{1}{c}{} &
  \multicolumn{1}{c}{(\AA)} &
  \multicolumn{1}{c}{(erg/cm$^2$/s/\AA)} &
  \multicolumn{1}{c}{(erg/cm$^2$/s/\AA)}  \\
\noalign{\smallskip}\hline
\noalign{\smallskip}
\endhead
\hline
\endfoot

% Start table
 
ChaI-PM-1	&     TYCHO/TYCHO.B	&    4280.0  &   3.7725337614949E-14  &   8.443355761978E-15   \\
ChaI-PM-1	& Generic/Johnson.B	&    4378.1  &   3.6617001607623E-14  &   1.0117651446222E-15  \\ 
ChaI-PM-1	&      SLOAN/SDSS.g	&    4640.4  &   5.72443206827E-14    &   1.0544793559003E-15  \\ 
ChaI-PM-1	&     TYCHO/TYCHO.V	&    5340.0  &   8.5480686453463E-14  &   8.266715283399E-15   \\
ChaI-PM-1	& Generic/Johnson.V	&    5466.1  &   7.7129282183686E-14  &   7.1038694155794E-16  \\ 
ChaI-PM-1	&      SLOAN/SDSS.r	&    6122.3  &   1.0060360371168E-13  &   9.265934328653E-16   \\
ChaI-PM-1	&      SLOAN/SDSS.i	&    7439.5  &   1.1753320341011E-13  &   5.4126040422267E-15  \\ 
ChaI-PM-1	&     DENIS/DENIS.I	&    7862.1  &   1.0792666849298E-13  &   6.9582894642363E-15  \\ 
ChaI-PM-1	&     2MASS/2MASS.J	&   12350.0  &   8.0947895646707E-14  &   1.7893384111233E-15  \\ 
ChaI-PM-1	&     2MASS/2MASS.H	&   16620.0  &   5.8267972939798E-14  &   2.4686721723789E-15  \\ 
  \multicolumn{5}{c}{...}\\
 
  \noalign{\smallskip}
  \hline
\end{longtable}

%\begin{flushleft}
%{\bf References:} \\
$^{\mathrm{a}}$ References:
    
    Johnson, Sloan: UCAC4/APASS, \cite{zacharias2013}
    
    Tycho-2: \citet{hog2000} 
    
    Stromgren: \citet{hauck1998} 
    
    DENIS: \citet{denis2005} 
    
    2MASS: \citet{skrutskie2006} 
    
    WISE: \citet{wright2010} 
    
    Spitzer: \citet{alcala2008,luhman2008c}; unpublished 
    
    AKARI: \citet{ishihara2010}
    
    IRAS: \citet{helou1988}
    
$^{\mathrm{b}}$ A null error value indicates a flux upper limit.
%\end{flushleft}

}

}
%----------------------------------------------------------

\longtabL{6}{
\begin{landscape}
\begin{longtable}{l l r r r r r r r r r r c c}

% first head
\caption{Best-fitting results and derived masses and ages for our kinematical candidate members to the Cha~I and II moving groups}
\label{tab:fitres}\\
    \hline \hline \noalign{\smallskip}
  \multicolumn{1}{l}{Object} &
  \multicolumn{1}{l}{Model} &
  \multicolumn{1}{r}{$\dot{\chi}^2$} &
  \multicolumn{1}{c}{$T_{eff}$} &
  \multicolumn{1}{c}{P($T_{eff}$)} &
  \multicolumn{1}{c}{$\log\,g$} &
  \multicolumn{1}{c}{P($\log\,g$)} &
  \multicolumn{1}{c}{$A_v$} &
  \multicolumn{1}{c}{P($A_v$)} &
  \multicolumn{1}{c}{$\log\,L/L_{\odot}$} &
  \multicolumn{1}{r}{Age} &
  \multicolumn{1}{r}{$M$} &
  \multicolumn{1}{c}{Classification$^{\mathrm{a}}$} &
  \multicolumn{1}{c}{Remarks$^{\mathrm{b}}$} \\
  \multicolumn{1}{l}{} &
  \multicolumn{1}{l}{} &
  \multicolumn{1}{r}{} &
  \multicolumn{1}{c}{(K)} &
  \multicolumn{1}{c}{} &
  \multicolumn{1}{c}{} &
  \multicolumn{1}{c}{} &
  \multicolumn{1}{c}{(mag)} &
  \multicolumn{1}{c}{} &
  \multicolumn{1}{c}{} &
  \multicolumn{1}{r}{(Myr)} &
  \multicolumn{1}{r}{($M_{\odot}$)} &
  \multicolumn{1}{c}{} &
  \multicolumn{1}{c}{} \\
      \noalign{\smallskip}\hline\noalign{\smallskip}
\endfirsthead

%second heads
\caption{(continued)}\\
   \hline\hline\noalign{\smallskip}
  \multicolumn{1}{l}{Object} &
  \multicolumn{1}{l}{Model} &
  \multicolumn{1}{r}{$\dot{\chi}^2$} &
  \multicolumn{1}{c}{$T_eff$} &
  \multicolumn{1}{c}{P($T_eff$)} &
  \multicolumn{1}{c}{$\log\,g$} &
  \multicolumn{1}{c}{P($\log\,g$)} &
  \multicolumn{1}{c}{$A_v$} &
  \multicolumn{1}{c}{P($A_v$)} &
  \multicolumn{1}{c}{$\log\,L/L_{\odot}$} &
  \multicolumn{1}{r}{Age} &
  \multicolumn{1}{r}{$M$} &
  \multicolumn{1}{c}{Classification$^{\mathrm{a}}$} &
  \multicolumn{1}{c}{Remarks$^{\mathrm{b}}$} \\
  \multicolumn{1}{l}{} &
  \multicolumn{1}{l}{} &
  \multicolumn{1}{r}{} &
  \multicolumn{1}{c}{(K)} &
  \multicolumn{1}{c}{} &
  \multicolumn{1}{c}{} &
  \multicolumn{1}{c}{} &
  \multicolumn{1}{c}{(mag)} &
  \multicolumn{1}{c}{} &
  \multicolumn{1}{c}{} &
  \multicolumn{1}{r}{(Myr)} &
  \multicolumn{1}{r}{($M_{\odot}$)} &
  \multicolumn{1}{c}{} &
  \multicolumn{1}{c}{} \\
      \noalign{\smallskip}\hline\noalign{\smallskip}
\endhead
\hline
\endfoot

% Start table
  \multicolumn{14}{c}{Cha~I moving group}\\
  \noalign{\smallskip}
  \object{ChaI-PM-1}  & BT-Settl &   8.85 & 4000 & 0.59 & 4.0 & 0.51 & 0.05 & 0.06 &	0.09$\pm$0.10 &  2 & 1.15 & m  & \\
  \object{ChaI-PM-2}  & BT-Settl &  12.91 & 4100 & 0.62 & 3.5 & 0.70 & 0.35 & 0.15 & $-$0.27$\pm$0.10 &    &	  & nc & \\
  \object{ChaI-PM-3}  & BT-Settl &   4.85 & 4200 & 0.67 & 4.0 & 0.47 & 0.00 & 1.00 & $-$0.12$\pm$0.11 &  7 & 1.2  & m  & \\
  \object{ChaI-PM-4}  & BT-Settl &   7.01 & 3900 & 0.76 & 4.0 & 0.50 & 0.12 & 0.06 &	0.34$\pm$0.10 &    &	  & m? & \\
  \object{ChaI-PM-5}  & BT-Settl &  19.39 & 4000 & 0.67 & 4.0 & 0.73 & 0.68 & 0.19 &	0.72$\pm$0.10 &    &	  & nc & \\
  \object{ChaI-PM-6}  & BT-Settl &   4.62 & 4000 & 0.97 & 4.0 & 0.70 & 0.00 & 1.00 &	0.08$\pm$0.10 &  2 & 1.15 & m  & \\
  \object{ChaI-PM-7}  & BT-Settl &   8.61 & 4100 & 0.96 & 3.5 & 0.54 & 0.00 & 1.00 & $-$0.42$\pm$0.10 & 14 & 1.0  & m  & \\
  \object{ChaI-PM-8}  & BT-Settl &   8.17 & 4100 & 0.93 & 3.5 & 0.47 & 0.00 & 1.00 & $-$0.19$\pm$0.10 &  7 & 1.1  & m  & \\
  \object{ChaI-PM-9}  & BT-Settl &  14.13 & 4100 & 0.94 & 3.5 & 0.36 & 0.00 & 1.00 & $-$0.23$\pm$0.10 &    &	  & nc & \\
  \object{ChaI-PM-11} & BT-Settl &   6.13 & 4200 & 0.47 & 4.0 & 0.48 & 0.09 & 0.06 & $-$0.05$\pm$0.10 &  5 & 1.3  & m  & \\
  \object{ChaI-PM-12} & BT-Settl &  13.23 & 3800 & 0.99 & 4.0 & 0.92 & 0.00 & 1.00 & $-$0.71$\pm$0.10 &    &	  & nc & \\
  \object{ChaI-PM-13} & BT-Settl &  13.48 & 4000 & 0.62 & 4.0 & 0.70 & 0.86 & 0.28 &	0.67$\pm$0.10 &    &	  & nc & \\
  \object{ChaI-PM-14} & BT-Settl &  44.62 & 4100 & 0.67 & 4.0 & 0.50 & 1.56 & 0.15 & $-$0.13$\pm$0.09 &    &	  & nc & \\
  \object{ChaI-PM-16} & BT-Settl &  19.00 & 3700 & 0.90 & 4.0 & 0.54 & 0.00 & 1.00 &	0.30$\pm$0.10 &    &	  & nc & \\
  \object{ChaI-PM-17} & BT-Settl &  86.03 & 2800 & 1.00 & 4.0 & 1.00 & 0.00 & 1.00 &	0.84$\pm$0.10 &    &	  & nc & \\ 
  \object{ChaI-PM-18} & BT-Settl & 104.39 & 3000 & 1.00 & 5.0 & 1.00 & 0.00 & 1.00 & $-$1.06$\pm$0.10 &    &	  & nc & \\ 
  \object{ChaI-PM-19} & BT-Settl &  18.85 & 4100 & 0.98 & 4.5 & 0.52 & 0.00 & 1.00 & $-$0.16$\pm$0.10 &    &	  & nc & \\
  \object{ChaI-PM-20} & BT-Settl & 111.33 & 2800 & 0.99 & 3.5 & 0.95 & 0.00 & 1.00 & $-$0.15$\pm$0.10 &    &	  & nc & \\ 
  \object{ChaI-PM-21} & BT-Settl &   9.06 & 4200 & 0.98 & 5.0 & 0.34 & 0.00 & 1.00 & $-$0.42$\pm$0.10 & 19 & 1.0  & m  & \\
  \object{ChaI-PM-22} & BT-Settl &   8.77 & 4000 & 0.94 & 4.0 & 0.44 & 0.00 & 1.00 & $-$0.11$\pm$0.10 &  4 & 1.1  & m  & \\
  \object{ChaI-PM-23} & BT-Settl &   9.26 & 4000 & 0.93 & 4.0 & 0.71 & 0.16 & 0.18 & $-$0.10$\pm$0.10 &  4 & 1.1  & m  & \\
  \object{ChaI-PM-24} & BT-Settl &   5.83 & 4200 & 0.67 & 4.5 & 0.32 & 0.00 & 1.00 & $-$0.13$\pm$0.14 &  7 & 1.2  & m  & \\
  \object{ChaI-PM-25} & BT-Settl &  16.68 & 3900 & 0.98 & 4.5 & 0.44 & 0.00 & 1.00 & $-$0.21$\pm$0.10 &    &	  & nc & \\
  \object{ChaI-PM-26} & BT-Settl &  11.51 & 4200 & 0.78 & 4.5 & 0.31 & 0.00 & 1.00 & $-$0.01$\pm$0.11 &    &	  & nc & \\
  \object{ChaI-PM-27} & BT-Settl &   4.84 & 4000 & 0.94 & 4.0 & 0.73 & 0.00 & 1.00 &	0.09$\pm$0.10 &  2 & 1.15 & m  & \\
  \object{ChaI-PM-28} & BT-Settl &  15.00 & 3200 & 0.68 & 5.0 & 0.29 & 0.00 & 1.00 & $-$0.79$\pm$0.10 &    &	  & nc & \\
  \object{ChaI-PM-29} & BT-Settl &  54.19 & 3400 & 0.96 & 3.5 & 0.95 & 0.00 & 1.00 & $-$0.20$\pm$0.10 &    &	  & nc & \\
  \object{ChaI-PM-30} & BT-Settl &   6.29 & 4100 & 0.97 & 4.5 & 0.49 & 0.00 & 1.00 & $-$0.55$\pm$0.10 & 20 & 0.9  & m  & \\
  \object{ChaI-PM-31} & BT-Settl &  24.35 & 3600 & 0.98 & 4.0 & 0.91 & 0.00 & 1.00 &	0.14$\pm$0.14 &    &	  & nc & \\
  \object{ChaI-PM-32} & BT-Settl &  17.44 & 3800 & 0.54 & 4.0 & 0.55 & 0.00 & 1.00 & $-$0.77$\pm$0.10 &    &	  & nc & \\
  \object{ChaI-PM-33} & BT-Settl &  14.79 & 3600 & 0.56 & 4.0 & 0.96 & 0.00 & 1.00 & $-$0.75$\pm$0.10 &    &	  & nc & \\
  \object{ChaI-PM-34} & BT-Settl &  19.06 & 4000 & 1.00 & 4.0 & 0.65 & 0.00 & 1.00 &	0.17$\pm$0.10 &    &	  & nc & \\
  \object{ChaI-PM-35} & BT-Settl &  13.11 & 3800 & 0.92 & 4.0 & 0.50 & 0.00 & 1.00 &	0.61$\pm$0.10 &    &	  & nc & \\
  \object{ChaI-PM-37} & BT-Settl &  43.45 & 3000 & 1.00 & 4.0 & 1.00 & 0.00 & 1.00 & $-$0.67$\pm$0.10 &    &	  & nc & \\
  \object{ChaI-PM-38} & BT-Settl &  49.70 & 3000 & 0.86 & 4.0 & 0.88 & 0.00 & 1.00 &	0.83$\pm$0.10 &    &	  & nc & \\ 
  \object{ChaI-PM-39} & BT-Settl &   4.67 & 4100 & 0.55 & 4.0 & 0.34 & 0.00 & 1.00 &	0.31$\pm$0.10 &  1 & 1.3  & m  & \\
  \object{ChaI-PM-40} & BT-Settl &  19.13 & 3700 & 0.97 & 4.5 & 0.41 & 0.00 & 1.00 & $-$0.11$\pm$0.10 &    &	  & nc & \\
  \object{ChaI-PM-41} & BT-Settl &  17.99 & 3800 & 1.00 & 3.5 & 0.50 & 0.00 & 1.00 &	0.59$\pm$0.10 &    &	  & nc & \\
  \object{ChaI-PM-42} & BT-Settl &  13.23 & 3600 & 0.64 & 4.0 & 0.87 & 0.00 & 1.00 & $-$0.65$\pm$0.10 &    &	  & nc & \\
  \object{ChaI-PM-43} & BT-Settl &  18.82 & 3800 & 1.00 & 3.5 & 0.97 & 0.00 & 1.00 & $-$0.58$\pm$0.10 &    &	  & nc & \\
  \object{ChaI-PM-45} & BT-Settl &   5.15 & 4200 & 0.71 & 4.0 & 0.51 & 0.00 & 1.00 & $-$0.32$\pm$0.10 & 12 & 1.1  & m  & \\
  \object{ChaI-PM-46} & BT-Settl &  12.44 & 4000 & 1.00 & 4.0 & 0.71 & 0.00 & 1.00 & $-$0.58$\pm$0.10 &    &	  & nc & \\
  \object{ChaI-PM-47} & BT-Settl &  44.63 & 3500 & 0.86 & 5.0 & 0.86 & 0.00 & 1.00 &	0.28$\pm$0.10 &    &	  & nc & \\
  \object{ChaI-PM-48} & BT-Settl &   3.13 & 3400 & 0.85 & 3.5 & 0.57 & 0.00 & 1.00 & $-$0.87$\pm$0.13 &  4 & 0.3  & m  & 2 \\
  \object{ChaI-PM-50} & BT-Settl &  59.24 & 3700 & 0.52 & 4.0 & 0.48 & 0.00 & 1.00 &	0.58$\pm$0.10 &    &	  & nc & \\
  \object{ChaI-PM-51} & BT-Settl &  12.53 & 3900 & 0.68 & 3.5 & 0.35 & 0.00 & 1.00 &	0.26$\pm$0.14 &    &	  & nc & \\
  \object{ChaI-PM-52} & BT-Settl &  16.16 & 3200 & 0.82 & 4.0 & 0.61 & 0.00 & 1.00 & $-$0.77$\pm$0.10 &    &	  & nc & \\
  \object{ChaI-PM-53} & BT-Settl &  10.83 & 4200 & 1.00 & 4.0 & 0.56 & 0.00 & 1.00 & $-$0.38$\pm$0.10 &    &	  & nc & \\
  \object{ChaI-PM-55} & BT-Settl &  21.71 & 4000 & 1.00 & 4.0 & 0.91 & 0.00 & 1.00 &	0.02$\pm$0.10 &    &	  & nc & \\
  \object{ChaI-PM-56} & BT-Settl &   9.97 & 4100 & 0.99 & 4.5 & 0.73 & 0.00 & 1.00 & $-$0.20$\pm$0.10 &  7 & 1.1  & m  & \\
  \object{ChaI-PM-57} & BT-Settl &  10.03 & 4200 & 0.79 & 3.5 & 0.42 & 0.00 & 1.00 & $-$0.42$\pm$0.14 & 20 & 1.0  & m  & \\

  \noalign{\smallskip}\hline\noalign{\smallskip}
  \multicolumn{14}{c}{Cha~II moving group}\\
  \noalign{\smallskip}

  \object{ChaII-PM-1}  & BT-Settl &  14.34 & 3600 & 0.31 & 5.0 & 0.84 & 1.69 & 0.20 & $-$0.61$\pm$0.10 &   &	 & nc & \\  
  \object{ChaII-PM-2}  & BT-Settl &  22.74 & 3800 & 0.87 & 4.0 & 0.88 & 0.92 & 0.82 &	 0.48$\pm$0.10 &   &	 & nc & \\
  \object{ChaII-PM-3}  & BT-Settl &  85.64 & 3600 & 0.93 & 4.0 & 0.78 & 2.80 & 0.85 &	 0.73$\pm$0.10 &   &	 & nc & \\
  \object{ChaII-PM-6}  & BT-Settl &  13.93 & 3800 & 0.39 & 4.0 & 0.48 & 1.18 & 0.33 & $-$0.08$\pm$0.10 &   &	 & nc & \\
  \object{ChaII-PM-7}  & BT-Settl &  37.13 & 3400 & 0.90 & 4.0 & 0.90 & 0.24 & 0.20 & $-$0.22$\pm$0.10 &   &	 & nc & \\  
  \object{ChaII-PM-8}  & BT-Settl & 100.79 & 3200 & 1.00 & 4.0 & 1.00 & 0.00 & 1.00 & $-$0.23$\pm$0.11 &   &	 & nc & \\
  \object{ChaII-PM-9}  & BT-Settl & 270.40 & 2800 & 1.00 & 5.0 & 1.00 & 0.00 & 1.00 &	 1.10$\pm$0.11 &   &	 & nc & \\  
  \object{ChaII-PM-10} & BT-Settl &  61.97 & 3400 & 1.00 & 4.0 & 1.00 & 0.00 & 1.00 &	 0.01$\pm$0.10 &   &	 & nc & \\  
  \object{ChaII-PM-11} & BT-Settl &   6.34 & 3300 & 0.64 & 3.5 & 0.55 & 0.00 & 1.00 & $-$0.74$\pm$0.10 & 2 & 0.3 & m  & 1 \\
  \object{ChaII-PM-12} & BT-Settl &  52.85 & 3600 & 1.00 & 4.0 & 1.00 & 0.00 & 1.00 &	 0.52$\pm$0.11 &   &	 & nc & \\
  \object{ChaII-PM-13} & BT-Settl &  35.37 & 2800 & 1.00 & 5.0 & 0.80 & 0.00 & 1.00 &	 0.41$\pm$0.11 &   &	 & nc & \\
  \object{ChaII-PM-14} & BT-Settl &   4.49 & 2800 & 1.00 & 5.0 & 0.86 & 0.00 & 1.00 &	 1.60$\pm$0.13 &   &	 & nm & \\ 
  \object{ChaII-PM-17} & BT-Settl &  55.21 & 3400 & 0.98 & 4.5 & 0.38 & 0.00 & 1.00 &	 0.37$\pm$0.11 &   &	 & nc & \\
  \object{ChaII-PM-18} & BT-Settl &  48.59 & 2900 & 1.00 & 4.5 & 0.56 & 0.00 & 1.00 &	 1.39$\pm$0.10 &   &	 & nm & \\

  \noalign{\smallskip}
  \hline
\end{longtable}

 \begin{flushleft}
 {\bf Notes.} \\
      $^{\mathrm{a}}$ m$=$probable member; m?$=$possible member; nm$=$nonmember; nc$=$not classified \\
      $^{\mathrm{b}}$
      (1)$=$Object with possible mid-infrared excess starting at $\sim10~\mu$m;
      (2)$=$Object with possible mid-infrared excess starting at $\sim20~\mu$m.
 \end{flushleft}

\end{landscape}
}

 % Cham I & II  longtable

%----------------------------------------------------------
\setcounter{table}{7}

%_____________________________________________________________
%                              Table longer than a single page
%                                             and in landscape 
%  In the preamble, use:       \usepackage{longtable,lscape}
%-------------------------------------------------------------
%          All long tables have to be placed at the end, after
%                                        \end{thebibliography}
%
% In the text, at the place where the large table should appear
% add the command:                        
%\addtocounter{table}{1} 
% Tables counters will be well numbered.
%-------------------------------------------------------------

\longtabL{8}{
\begin{landscape}
\begin{longtable}{l l r c c r c c c r c c c}

% first head

\caption{Physical parameters of Chamaeleon~I and II members} \label{tab:mempars}\\
    \hline 
    \hline 
    \noalign{\smallskip}
  \multicolumn{1}{c}{2MASS~J} &
  \multicolumn{1}{c}{Model} &
  \multicolumn{1}{c}{$\dot{\chi}^2$} &
  \multicolumn{1}{c}{$T_{eff}$} &
  \multicolumn{1}{c}{P($T_{eff}$)} &
  \multicolumn{1}{c}{$\log\,g$} &
  \multicolumn{1}{c}{P($\log\,g$)} &
  \multicolumn{1}{c}{$A_V$} &
  \multicolumn{1}{c}{P($A_V$)} &
  \multicolumn{1}{c}{$\log(L/L_{\odot})$} &
  \multicolumn{1}{c}{Age} &
  \multicolumn{1}{c}{$M$} &
  \multicolumn{1}{c}{Class$^{\mathrm{a}}$} \\
  \multicolumn{1}{c}{} &
  \multicolumn{1}{c}{} &
  \multicolumn{1}{c}{} &
  \multicolumn{1}{c}{(K)} &
  \multicolumn{1}{c}{} &
  \multicolumn{1}{c}{} &
  \multicolumn{1}{c}{(mag)} &
  \multicolumn{1}{c}{} &
  \multicolumn{1}{c}{} &
  \multicolumn{1}{c}{} &
  \multicolumn{1}{c}{(Myr)} &
  \multicolumn{1}{c}{($M_{\odot}$)} &
  \multicolumn{1}{c}{} \\
\noalign{\smallskip}\hline\noalign{\smallskip}
\endfirsthead

%second heads
\caption{(continued)}\\
    \hline 
    \hline 
    \noalign{\smallskip}
  \multicolumn{1}{c}{2MASS~J} &
  \multicolumn{1}{c}{Model} &
  \multicolumn{1}{c}{$\dot{\chi}^2$} &
  \multicolumn{1}{c}{$T_{eff}$} &
  \multicolumn{1}{c}{P($T_{eff}$)} &
  \multicolumn{1}{c}{$\log\,g$} &
  \multicolumn{1}{c}{P($\log\,g$)} &
  \multicolumn{1}{c}{$A_V$} &
  \multicolumn{1}{c}{P($A_V$)} &
  \multicolumn{1}{c}{$\log(L/L_{\odot})$} &
  \multicolumn{1}{c}{Age} &
  \multicolumn{1}{c}{$M$} &
  \multicolumn{1}{c}{Class$^{\mathrm{a}}$} \\
  \multicolumn{1}{c}{} &
  \multicolumn{1}{c}{} &
  \multicolumn{1}{c}{} &
  \multicolumn{1}{c}{(K)} &
  \multicolumn{1}{c}{} &
  \multicolumn{1}{c}{} &
  \multicolumn{1}{c}{(mag)} &
  \multicolumn{1}{c}{} &
  \multicolumn{1}{c}{} &
  \multicolumn{1}{c}{} &
  \multicolumn{1}{c}{(Myr)} &
  \multicolumn{1}{c}{($M_{\odot}$)} &
  \multicolumn{1}{c}{} \\
\noalign{\smallskip}\hline
\noalign{\smallskip}
\endhead
\hline
\endfoot

% Start table
  \multicolumn{13}{c}{Chamaeleon~I cloud}\\
  \noalign{\smallskip}
  \object{10463795-7736035} & Kurucz   & 17.19 &  9250 & 0.20 & 3.5 & 0.35 & 2.0 & 0.63 &      1.41$\pm$0.09 &    &	      & II  \\
  \object{10574219-7659356} & BT-Settl & 46.33 &  5000 & 0.22 & 3.5 & 0.34 & 3.5 & 0.40 &      0.52$\pm$0.09 &    &	      & II  \\
  \object{11022610-7502407} & Kurucz   &  8.91 &  3000 & 1.00 & 3.5 & 1.00 & 0.0 & 1.00 &    --1.13$\pm$0.09 &  1 & 0.10:0.11 & II  \\
  \object{11023265-7729129} & Kurucz   &  1.83 & 24000 & 0.28 & 3.5 & 0.34 & 7.0 & 0.82 &      1.93$\pm$0.09 &    &	      & II  \\
  \object{11025504-7721508} & BT-Settl & 52.03 & 13000 & 0.06 & 3.5 & 0.35 & 6.4 & 0.37 &      2.81$\pm$0.09 &    &	      & II  \\
  \object{11035682-7721329} & BT-Settl &  8.26 &  3400 & 0.49 & 4.0 & 0.52 & 1.2 & 0.84 &    --0.65$\pm$0.09 &  2 & 0.38      & III \\
  \object{11045100-7625240} & BT-Settl &  3.29 &  3800 & 0.25 & 4.0 & 0.50 & 0.0 & 0.62 &    --0.59$\pm$0.12 & 10 & 0.80      & II  \\
  \object{11045285-7625514} & BT-Settl &  1.09 &  3600 & 0.35 & 4.0 & 0.52 & 0.0 & 0.98 &    --0.66$\pm$0.13 &  6 & 0.57      & II  \\
  \object{11050752-7812063} & BT-Settl & 97.75 &  2700 & 1.00 & 4.0 & 1.00 & 0.0 & 1.00 &    --1.17$\pm$0.09 &    &	      & II  \\
  \object{11051467-7711290} & BT-Settl &  5.75 &  3400 & 0.96 & 4.0 & 0.98 & 2.3 & 0.99 &    --0.52$\pm$0.09 &  2 & 0.40      & II  \\
  \object{11052472-7626209} & BT-Settl &  4.13 &  3300 & 0.39 & 3.5 & 0.71 & 0.0 & 0.77 &    --0.98$\pm$0.09 &  4 & 0.27      & III \\
  \object{11055261-7618255} & BT-Settl &  2.02 &  3500 & 0.33 & 4.0 & 0.61 & 0.0 & 0.98 &    --0.53$\pm$0.12 &  3 & 0.49      & II  \\
  \object{11061540-7721567} & BT-Settl &  8.78 &  3600 & 0.29 & 4.0 & 0.65 & 0.6 & 0.47 &      0.72$\pm$0.09 &    &	      & II  \\
  \object{11064346-7726343} & BT-Settl &  3.79 &  3400 & 0.27 & 3.5 & 0.46 & 2.3 & 0.43 &    --0.43$\pm$0.09 &  1 & 0.43      & TO  \\
  \object{11064510-7727023} & BT-Settl & 91.02 &  4100 & 0.64 & 4.5 & 0.91 & 2.3 & 0.65 &    --0.05$\pm$0.09 &  2 & 1.20      & II  \\
  \object{11065906-7718535} & Kurucz   & 37.32 &  7500 & 0.02 & 4.0 & 0.39 & 5.9 & 0.29 &      4.11$\pm$0.09 & 54 & 6.17      & II  \\
  \object{11071148-7746394} & BT-Settl &  4.58 &  3600 & 0.46 & 4.0 & 0.58 & 2.9 & 0.62 &    --0.48$\pm$0.09 &  3 & 0.60      & III \\
  \object{11071915-7603048} & BT-Settl &  3.46 &  3600 & 0.60 & 4.0 & 0.51 & 1.0 & 0.66 &    --0.63$\pm$0.09 &  5 & 0.59      & II  \\
  \object{11072040-7729403} & BT-Settl &  5.10 &  3100 & 0.50 & 3.5 & 0.50 & 0.0 & 0.76 &    --0.89$\pm$0.09 &    &	      & TO  \\
  \object{11072074-7738073} & BT-Settl & 58.71 &  3800 & 0.16 & 4.0 & 0.53 & 1.8 & 0.47 &      0.93$\pm$0.09 &    &	      & II  \\
  \object{11075588-7727257} & BT-Settl & 29.44 &  4500 & 0.22 & 3.5 & 0.48 & 3.5 & 0.43 &      0.54$\pm$0.09 &    &	      & II  \\
  \object{11080148-7742288} & BT-Settl & 64.09 &  4200 & 0.32 & 4.0 & 0.57 & 2.9 & 0.58 &      0.60$\pm$0.09 &    &	      & II  \\
  \object{11081648-7744371} & BT-Settl & 14.34 &  3600 & 0.96 & 4.0 & 0.80 & 1.8 & 0.99 &    --0.67$\pm$0.09 &  6 & 0.57      & II  \\
  \object{11084069-7636078} & BT-Settl & 45.30 &  3500 & 1.00 & 4.5 & 1.00 & 2.0 & 1.00 &    --0.36$\pm$0.09 &  2 & 0.53      & II  \\
  \object{11085464-7702129} & BT-Settl & 54.11 &  3600 & 0.39 & 4.0 & 0.88 & 2.9 & 0.57 &    --0.27$\pm$0.09 &  2 & 0.64      & II  \\
  \object{11094006-7628391} & BT-Settl &  5.18 &  3800 & 0.65 & 4.0 & 0.54 & 1.0 & 0.65 &    --0.28$\pm$0.09 &  4 & 0.90      & II  \\
  \object{11095003-7636476} & Kurucz   & 98.77 &  9250 & 0.45 & 3.5 & 0.76 & 2.0 & 1.00 &      1.55$\pm$0.09 &    &	      & II  \\
  \object{11095407-7629253} & BT-Settl & 64.16 &  4300 & 1.00 & 4.5 & 0.91 & 6.0 & 1.00 &    --0.07$\pm$0.09 &  7 & 1.30      & II  \\
  \object{11103801-7732399} & Kurucz   & 15.93 & 18000 & 0.23 & 3.5 & 0.39 & 7.6 & 0.62 &      2.35$\pm$0.09 &    &	      & II  \\
  \object{11105333-7634319} & BT-Settl & 40.21 &  4500 & 1.00 & 4.5 & 0.43 & 3.0 & 1.00 &    --0.51$\pm$0.09 & 40 & 0.85      & II  \\ 
  \object{11113474-7636211} & BT-Settl & 43.84 &  4200 & 0.98 & 4.0 & 0.99 & 2.0 & 1.00 &    --0.47$\pm$0.09 & 20 & 0.95      & II  \\
  \object{11124210-7658400} & Kurucz   &  7.24 &  5500 & 0.18 & 4.5 & 0.37 & 2.3 & 0.35 &    --0.03$\pm$0.09 &    &	      & III \\
  \object{11124268-7722230} & Kurucz   & 72.19 &  6250 & 0.85 & 4.0 & 0.41 & 2.3 & 0.85 &      0.88$\pm$0.09 &    &	      & TO  \\
  \object{11124299-7637049} & BT-Settl &  2.50 &  4200 & 0.52 & 4.5 & 0.49 & 0.0 & 1.00 &    --0.34$\pm$0.09 & 13 & 1.05      & TO  \\
  \object{11132012-7701044} & Kurucz   &  4.00 &  6750 & 0.10 & 4.5 & 0.46 & 3.5 & 0.42 &      0.02$\pm$0.09 &    &	      & TO  \\
  \object{11132446-7629227} & BT-Settl & 26.68 &  3100 & 0.87 & 3.5 & 0.87 & 0.0 & 0.87 &    --1.15$\pm$0.09 &  2 & 0.15      & II  \\
  \object{11132737-7634165} & BT-Settl &  7.20 &  3400 & 0.42 & 3.5 & 0.53 & 0.0 & 0.99 &    --0.64$\pm$0.09 &  2 & 0.38      & II  \\
  \object{11132970-7629012} & BT-Settl &  8.65 &  3000 & 0.99 & 3.5 & 0.99 & 0.0 & 0.99 &    --1.08$\pm$0.10 &    &	      & II  \\
  \object{11145031-7733390} & BT-Settl &  1.93 &  3400 & 0.47 & 4.0 & 0.54 & 0.0 & 0.60 &    --0.61$\pm$0.09 &  2 & 0.39      & TO  \\
  \object{11155827-7729046} & Kurucz   & 39.49 &  6750 & 0.15 & 3.5 & 0.35 & 4.1 & 0.37 &    --0.16$\pm$0.09 &    &	      & TO  \\
  \object{11242980-7554237} & BT-Settl & 10.00 &  3200 & 0.99 & 4.0 & 0.99 & 1.0 & 1.00 &    --0.71$\pm$0.09 &  1 & 0.20      & II  \\

  \noalign{\smallskip}\hline\noalign{\smallskip}
  \multicolumn{13}{c}{Chamaeleon~II cloud}\\
  \noalign{\smallskip}
  \object{12571172-7640111} & BT-Settl & 72.79 & 4200.0 & 0.98 & 4.0 & 0.89 & 4.0 & 0.99 &   0.25$\pm$0.10 &  2 & 1.40 & II  \\
  \object{13005323-7654151} & BT-Settl & 22.87 & 4000.0 & 0.99 & 4.5 & 0.87 & 2.0 & 0.99 & --0.82$\pm$0.10 & 40 & 0.75 & II  \\
  \object{13005532-7710222} & BT-Settl & 85.29 & 3400.0 & 0.50 & 4.0 & 0.97 & 3.0 & 0.52 & --0.03$\pm$0.10 &	&      & II  \\
  \object{13005534-7708296} & BT-Settl & 52.51 & 4000.0 & 0.93 & 4.0 & 0.94 & 3.5 & 0.98 & --0.36$\pm$0.10 &  9 & 1.00 & II  \\
  \object{13005622-7654021} & BT-Settl & 37.41 & 3900.0 & 0.86 & 3.5 & 0.98 & 0.5 & 0.86 &   0.31$\pm$0.10 &	&      & II  \\
  \object{13015891-7751218} & BT-Settl & 37.19 & 3400.0 & 0.80 & 4.0 & 0.86 & 1.5 & 0.80 & --0.23$\pm$0.10 &	&      & II  \\
  \object{13030444-7707027} & BT-Settl & 18.23 & 4900.0 & 0.78 & 3.5 & 0.85 & 2.5 & 0.83 &   0.10$\pm$0.10 &	& 1.24 & III \\
  \object{13045571-7739495} & BT-Settl & 44.56 & 4700.0 & 0.53 & 3.5 & 0.50 & 3.5 & 0.64 &   0.20$\pm$0.10 &	& 1.40 & II  \\
  \object{13052072-7739015} & BT-Settl & 34.85 & 3600.0 & 0.87 & 4.0 & 0.93 & 1.5 & 0.87 &   0.33$\pm$0.10 &	&      & II  \\
  \object{13063053-7734001} & BT-Settl & 36.11 & 3400.0 & 0.89 & 4.0 & 0.95 & 3.0 & 0.89 & --0.94$\pm$0.10 &  5 & 0.35 & II  \\
  \object{13072241-7737225} & BT-Settl & 11.57 & 3600.0 & 0.17 & 4.0 & 0.55 & 3.5 & 0.31 & --0.34$\pm$0.10 &  2 & 0.62 & II  \\
  \object{13074851-7741214} & BT-Settl & 11.63 & 3300.0 & 0.80 & 4.5 & 0.85 & 4.0 & 0.90 & --0.53$\pm$0.10 &  1 &      & II  \\
  \object{13095036-7757240} & BT-Settl & 43.81 & 4500.0 & 0.87 & 3.5 & 0.86 & 3.5 & 0.95 & --0.05$\pm$0.10 & 10 & 1.30 & II  \\
  \object{13100415-7710447} & BT-Settl & 22.65 & 3700.0 & 0.91 & 4.0 & 0.56 & 1.5 & 0.97 & --0.68$\pm$0.11 &  9 & 0.69 & II  \\
  \noalign{\smallskip}
  \hline
\end{longtable}

\begin{flushleft}
{\bf Notes.} \\
     $^{\mathrm{a}}$ II$=$class II source; III$=$class III source; TO$=$transitional object
\end{flushleft}

\end{landscape}
}

%----------------------------------------------------------

\end{document}